\documentclass[acmsmall]{acmart}
\usepackage{type1cm} 
\usepackage{graphicx} 
\usepackage{xspace} 
\usepackage{balance} 
\usepackage{booktabs} 
\usepackage{multirow} 
\usepackage[font={bf}, tableposition=top]{caption} 
\usepackage{bold-extra} 
\usepackage{siunitx} 
\usepackage[vlined,linesnumbered,ruled,noend]{algorithm2e} 
\usepackage{microtype} 
\usepackage{xfrac} 
\usepackage{mathtools} 
\PassOptionsToPackage{hyphens}{url} 
\PassOptionsToPackage{bookmarks, pdftex, colorlinks=true, pagebackref=true, backref=page}{hyperref} 
\usepackage{cleveref} 
\PassOptionsToPackage{square,numbers}{natbib} 
\usepackage[hyperpageref]{backref} 
\usepackage[htt]{hyphenat} 
\usepackage{subcaption} 
\usepackage[export]{adjustbox} 

\newcommand{\spara}[1]{\smallskip\noindent\textbf{#1}}

\renewcommand*\backref[1]{\ifx#1\relax \else (Cited on #1) \fi}

\newcommand{\ukr}{\texttt{@Ukraine}\xspace}
\newcommand{\umf}{\texttt{@uamemesforces}\xspace}
\newcommand{\dou}{\texttt{@DefenceU}\xspace}

\newcommand{\datasetUrl}{\url{https://github.com/corradomonti/ukranian-tweets}\xspace}

\newcommand{\rev}[1]{#1} 
\newcommand{\rvf}[1]{#1}

\graphicspath{{img/}}
\newcommand{\IncreaseNarrativeFool}{36\%\xspace}
\newcommand{\IncreaseNarrativeHero}{40\%\xspace}
\newcommand{\IncreaseNarrativeVictim}{109\%\xspace}
\newcommand{\IncreaseNarrativeVillain}{32\%\xspace}

\title{Narratives of War: Ukrainian Memetic Warfare on Twitter}

\author{Yelena Mejova}
\orcid{0000-0001-5560-4109}
\affiliation{%
  \institution{ISI Foundation}
  \city{Turin}
  \country{Italy}
}
\email{yelenamejova@acm.org}

\author{Arthur Capozzi}
\orcid{0000-0002-1996-1800}
\affiliation{%
  \institution{ETH Zürich}
  \city{Zürich}
  \country{Switzerland}
}
\email{arthur.capozzi@gess.ethz.ch}

\author{Corrado Monti}
\orcid{0000-0001-6846-5718}
\affiliation{%
  \institution{CENTAI}
  \city{Turin}
  \country{Italy}
}
\email{corrado.monti@centai.eu}

\author{Gianmarco De Francisci Morales}
\orcid{0000-0002-2415-494X}
\affiliation{%
  \institution{CENTAI}
  \city{Turin}
  \country{Italy}
}
\email{gdfm@acm.org}

\setcopyright{acmlicensed}
\acmJournal{PACMHCI}
\acmYear{2025} \acmVolume{9} \acmNumber{2} \acmArticle{CSCW139} \acmMonth{4}\acmDOI{10.1145/3711037}

\received{January 2024}
\received[revised]{July 2024}
\received[accepted]{October 2024}

\keywords{Memes, Propaganda, Social media}

\ccsdesc[500]{Information systems~Social networks}
\ccsdesc[500]{Human-centered computing~Empirical studies in HCI}
\ccsdesc[300]{Human-centered computing~Empirical studies in collaborative and social computing}

\begin{abstract}

The 2022 Russian invasion of Ukraine has seen an intensification in the use of social media by governmental actors in cyber warfare.
Wartime communication via memes has been a successful strategy used not only by independent accounts such as \umf, but also---for the first time in a full-scale interstate war---by official Ukrainian government accounts such as \ukr and \dou.
We study this prominent example of memetic warfare through the lens of its narratives, and find them to be a key component of success:
tweets with a `victim' narrative garner twice as many retweets.
However, malevolent narratives focusing on the enemy resonate more than those about heroism or victims with countries providing more assistance to Ukraine.
Our findings present a nuanced examination of Ukraine's influence operations and of the worldwide response to it, thus contributing new insights into the evolution of socio-technical systems in times of war.
\end{abstract}

\begin{document}

\maketitle

\section{Introduction}
\label{sec:intro}

On 24 February 2022, Russia began a large-scale military invasion of Ukraine.
The United Nations Human Rights Office has confirmed the death of \num{27768} civilians up to 8 October 2023.\footnote{\url{https://ukraine.un.org/en/248799-ukraine-civilian-casualties-8-october-2023}}
However, their report highlights that the actual figures are believed to be considerably higher, and most estimates suggest more than \num{300000} total deaths.

Amid this immense tragedy, the two governments at war---Russia and Ukraine---are also waging an information war on many fronts.
Due to the prevalence of the Internet as a primary information source for Ukrainians,\footnote{\url{https://www.unian.info/society/media-space-internet-bypasses-tv-as-main-news-source-for-ukrainians-11330729.html}} the war has been described as the first social-media war,\footnote{\url{https://www.forbes.com/sites/petersuciu/2022/03/01/is-russias-invasion-of-ukraine-the-first-social-media-war}} echoing the reputation of the Vietnam war as the ``first television war'' due to the role this medium had in informing the public about the conflict.

Both governments and the common people have been using social media to propagate their views on the conflict and to affect global public opinion.
A vivid example is \ukr, the official account of the Ukrainian government on Twitter.\footnote{Twitter has been rebranded to ``X''. For consistency with previous literature, we prefer using the former name.}
This account was started in 2016 by Yarema Dukh while working as a press officer in the Ukrainian presidential office.
One of the memes produced by this account, depicting ``living next to Russia'' as the worst possible headache,\footnote{\url{https://twitter.com/Ukraine/status/1468206078940823554}} has been shared more than \num{600000} times and has reached 55 million views.
According to the author, the goal of this meme was ``to explain to some large and distant target audiences that Russia is the problem here, not Ukraine, the West, the U.S., NATO, aliens or anyone else''.\footnote{\url{https://www.washingtonpost.com/world/2022/01/26/ukraine-russia-memes-social}}
This account %
has now become one of the most visible examples of memetic warfare from a warring government.

An early definition of memetic warfare was given by the NATO Strategic Communications Centre of Excellence in Latvia in a document titled \emph{``It's time to embrace memetic warfare''}~\cite{giesea2015s}.
The document defines it as ``the competition over narrative, ideas, and social control in a social-media battlefield'' and is viewed as ``the digital native's version of psychological warfare, more commonly known as propaganda''.
The authors argue that it needs to be brought into mainstream military thinking by investing in its development.
While propaganda has long been a central component of war, memetic warfare introduces a new key aspect: its reach is intrinsically tied to how much its audience is willing to spread it.
Governments engaged in psychological warfare used to rely \rev{mostly} on one-way, broadcast communication media such as posters and radio\rev{, and on subsequent word-of-mouth amplification by individuals}.
However, the two-way nature of social media implies that authors have to persuade their own audience to help spread their message \rev{on the platform} in order to reach wider audiences.
This characteristic creates a particular amalgam of social-media bottom-up dynamics with the top-down approach typical of military objectives and narratives.
Different objectives may then require spreading different types of content, which in turn may have varying degrees of success with specific audiences.

Despite the importance of this issue, so far little research effort has been dedicated to the quantitative analysis of \rvf{the effectiveness of memetic warfare. The goal of the present work is to investigate which audiences it is able to reach and resonate with.}
These questions are of primary interest not only to the parties involved, but also to global public opinion.
In order to prepare, evaluate, and react appropriately, we need to better understand these novel techniques of \rev{``soft propaganda''~\cite{mattingly2022how}}, which are likely to become more common in the future.

To concretely and critically evaluate these phenomena, we choose to analyze the main Ukrainian Twitter accounts involved in the Russo-Ukrainian conflict: the two main official Ukrainian government accounts, \ukr and \dou, and the most followed Ukrainian meme page, \umf.
Our analysis focuses on the following research questions:
\begin{itemize}
\item[RQ1] What are the main characteristics of the messaging used by these popular Ukrainian accounts?
\item[RQ2] Which attributes of a wartime meme by these popular Ukrainian accounts are good predictors of its virality?
\end{itemize}

\rev{Based on previous literature on memes, we hypothesize that the accounts will employ different visual elements~\cite{ling2021dissecting}. }
\rev{However, expanding on this literature,} we examine the content of memes in terms of the actors involved, their emotional appeal, and the narratives used.
\rev{Inspired by the narrative psychology and linguistic types analysis}~\cite{de2021internet}, we identify four roles within \rev{the memes}, depending on the moral quality of the actors in the narrative and on their perceived power.
In our context, these roles often represent the benevolent moral quality of the author's side or the malevolent one of the enemy\rev{, and we hypothesize that invocation of such qualities will result in greater popularity of the content}.
\rev{Indeed, we find that} the narratives we identify are a powerful predictor of the virality of a meme authored by these accounts.
Content with a victim narrative---a benevolent, weak actor---has a \IncreaseNarrativeVictim higher reach w.r.t. content with no narrative, while a villain narrative---depicting the enemy as malevolent and strong---increases reach by \IncreaseNarrativeVillain.

Finally, since one of the goals of memetic warfare is to influence global public opinion, we look at the success of the content geographically.
Considering the importance of the narratives we identify, we ask ourselves the following research question:
\begin{itemize}
\item[RQ3] What is the geographic resonance of each narrative, and how does it relate to each country's actions in the conflict?
\end{itemize}

\rev{As the Web platforms have been employed to rally support for war efforts~\cite{ye2023online}, we hypothesize that it is possible that the broader affinity to a narrative would be reflected in a country's willingness to assist one of the sides.}
We find that memetic warfare reach reflects important aspects of the offline world: the number of retweets per capita in a given country is highly correlated ($\rho=0.787$, $p<10^{-3}$) with the assistance given by that country to Ukraine.
In addition, different narratives resonate differently with different countries.
Tweets with a villain narrative spread more effectively in countries that are more involved in the conflict---as measured by the share of GDP devoted to economic and military assistance to Ukraine ($\rho = 0.573$, $p < 0.01$).
This connection emerges as significant not only when considering our data set as a whole, but even when analyzing each account individually, thus hinting at a more general pattern that could characterize communication about the Ukrainian war on social media.

The present study has important implications for our understanding of the interaction between communication technologies and their use by both governments and non-governmental actors during a conflict.
\rvf{Our research questions (particularly RQ2 and RQ3) focus on analyzing how the global audience perceives these messages.
In doing so,} we illustrate how social media can contribute to a country's strategic messaging during war, and how such messaging is related to the attention and aid given to the country. 
As cyber-influence campaigns become more common and sophisticated, we encourage the research community to apply a multi-factorial view of the relevant content, including visual and narrative aspects, to understand its resonance worldwide. 

\section{Background and Related Work}
\label{sec:relwork}

The social computing community has long been invested in exposing the horrors of war~\cite{hourcade2011chi, cannanure2023chi}, for example by studying how social media exacerbate mass trauma~\cite{scott2023trauma}. 
The community has analyzed how computer-mediated communication impacts politically-sensitive topics, e.g., by revealing the role of demagoguery in civic engagement on Reddit~\cite{Papakyriakopoulos2023upvotes}, participating in online activism (``slacktivism'') in donation campaigns~\cite{lee2013does}, and investigating the reach of anti-migration ad targeting on Facebook~\cite{capozzi2021clandestino}. 
Further, recent work on social media communication by politicians during crises has shown that their content receives outsized engagement that is not accounted for by their popularity or engaging tweet features~\cite{glunt2019public}---thus motivating further study on the role of official accounts.

The 2022 Russian invasion of Ukraine happened at a time of wide adoption of social media, when the advancement of Internet infrastructure enables real-time, high-resolution media to be shared with a worldwide audience.
As such, the actions on the battlefield and the reactions of those inside and outside Ukraine are streamed on Facebook, Twitter, Instagram, and other social media platforms.
This constant stream also ``proves to be an essential mechanism in getting on-ground real-time reporting of a dangerous event'' for mainstream media sources~\cite{suciu2022is}.
Several theoretical frameworks can help us frame such presence.
First, as a \emph{cyber-influence campaign}, which seeks to promote a narrative alternative to that of the Russian side~\cite{johnson2020semantically}. 
The second framework is that of \emph{nation-building} or \emph{nation-branding}: two processes that were originally separated in the domains of the government (former) and PR campaign managers (latter), %
but that have coalesced into a single process.
\citet{bolin2023nation} observe that the recent campaigns by Ukraine to promote ``Ukrainian capability'' are ``aimed to raise support for EU membership, arms deliveries''%
---a kind of soft power dubbed \emph{selfie diplomacy}~\cite{manor2015america}.
The third framework is that of \emph{community building} during a crisis that ``facilitates stakeholder support and builds relationship''~\cite{du2018social,coombs2021ongoing},
and targets both international and domestic audiences.

Since the annexation of Crimea in 2014, the social media presence of both Ukraine and Russia has been closely scrutinized by researchers in communications, media studies, and human-computer interaction.
The separatist movement in the Donbass region has intensified a battle of narratives:
both sides employed five contextual frames, namely historical, geographical, religious, ethnic, and political~\cite{makhortykh2015savedonbasspeople}.
According to \citet{makhortykh2017social}, those on the pro-Russian side were also more likely to make extensive use of photos of children, the authors speculate, in order to ``evoke compassion from the potential audience by using sentimental images''. 
On the Russian social media VKontakte in 2014, pro-Ukrainian users framed the conflict as a limited military action against local insurgents, while pro-Russian separatists framed it as an ``all-out war against the Russian population of Eastern Ukraine''~\citep{makhortykh2017social}.
Such divergent frames might have ``led to the formation of divergent expectations in Ukraine and Russia concerning the outcome of the war in Donbass''~\cite{makhortykh2017social}.

After the full-scale Russian invasion, several datasets have been made available to the research community. 
For instance,~\citet{chen2023tweets} collected Twitter posts having various Russo-Ukrainian, war-related keywords.
Focusing on the accounts participating in these online conversations, 
\citet{hare2023slava} track the use of the Ukrainian flag as a marker of support on Twitter in late February 2022---a form of identity activism~\cite{hare2023slava}.
Such accounts are homophilic, %
and more likely to share U.S.~Democrat-leaning messages.
Automated accounts, or bots, are often part of a communication operation.
\citet{shen2023examining} estimate the share of bot accounts around the beginning of Russia's invasion in 2022. 
Using the tool Botometer~\cite{yang2022botometer}, they identify around 13.4\% of the tweets as likely to be generated by bots.
Most of these tweets espoused a pro-Ukrainian position; it is worth noting that by then Russia had suspended access to Twitter for its citizens.

The term \emph{meme} was originally defined by the evolutionary biologist Richard Dawkins in 1976~\cite{dawkins1976selfish} as a replicator analogous to a gene in its ability to transmit information, including cultural artifacts and beliefs. %
Due to the vague nature of the term's definition, and the fact that the very nature of memes is a ``continuous mutation'', the study of memes, or Memetics, is applied broadly to the study of cultural information transfer~\cite{pittphilsci9912}.
In the age of social media, the concept of ``Internet meme'' breaks from Dawkin's analogy to focus on the artifacts---texts, images, or videos---instead of abstract ideas~\cite{lee2020neo}. 
Indeed, \citet{reese2001prologue} point to the properties of the visual medium, including syntactic implicitness and iconicity, that make it particularly suited for ``framing and articulating ideological messages''.
As memes are defined by their capability of being modified and reshared, we consider all visual media as potential memes~\cite{zenner2018one}.
Their virality potential has been analyzed using a set of visual features---e.g., scale or inclusion of text---by ~\citet{ling2021dissecting}: we use such features as a baseline in our analysis.
The usage of memes in political communication
have been studied during the 2016 US Presidential election~\cite{woods2019make}, the 2019 Ukrainian election~\cite{pidkuuimukha2020battle}, in Brazil~\cite{daSilva2021memes} and Hong Kong~\cite{dynel2021caveat}, in Germany's~\cite{bogerts2019you} and US'~\cite{dafaure2020great,greene2019deplorable} far right. %
Military actions in the past decade have been accompanied by meme-supported cultural expression by U.S. troops~\cite{silvestri2016mortars}, by anti-Islamic State activists~\cite{mccrow2021countering}, and in the early days of the Russo-Ukrainian conflict around Crimea~\cite{wiggins2016crimea}.
Finally, a notable precedent in the usage of memes by state military as a tool of propaganda is that of the Israel Defense Forces~\cite{massa2022platformization}.

\rev{The use of internet memes as a propaganda tool both echoes classic communication strategies, and employs new technologies for a greater impact.
Following past tactics, emotional, divisive messages are used to demonize, discredit, and ``other'' the opposition~\cite{woods2019make,dafaure2020great}.
Unlike the past top-down communication, current communication platforms host both official and user-generated content.
This content may be designed to look funny, simplistic and amateurish to be perceived as belonging to a grassroots movement instead of a coordinated communication campaign~\cite{daSilva2021memes}.
When the messages are forwarded by their audience, the easily accessible editing technologies allow for their content to be quickly modified and remixed, allowing for a ``participatory digital culture'' that is diverse and highly personalized~\cite{wiggins2016crimea}.
For instance, the hacker collective Anonymous~\cite{mccrow2021countering}, personnel in the US army~\cite{silvestri2016mortars}, or individuals bonded by a political view~\cite{dafaure2020great} may now produce and propagate politically-salient messages.
Moreover, the fractured social Web sphere creates spaces (including highly polarized and radicalized ones) wherein memes originate, develop, propagate, and eventually influence other communities and the mainstream media~\cite{zannettou2018origins}. 
Such content is often rife with inside culture references, humor, and attention-grabbing imagery~\cite{mccrow2021countering,pidkuuimukha2020battle,dynel2021caveat}, which may then be imitated by state actors~\cite{zannettou2020characterizing}}.

A popular framework for the analysis of memes relies on identifying the overarching \emph{narratives}, whereby each meme constitutes a partial element~\cite{de2021internet}.
\citet{boatwright2023don} study 163 tweets by the verified @Ukraine and @Kyiv Twitter accounts posted in February 2022,
finding pre-invasion promotional content as well as myth-making.
Similarly, \citet{yehorova2023towards} identify different humorous elements in the early wartime tweets.
A more structured framework of narrative analysis is that proposed by Stephen Karpman~\cite{karpman1968fairy} and expanded by \citet{de2021internet} to include two dimensions: 
moral quality (benevolent or malevolent) and power (strong or weak).
In our study, we employ this framework and the resulting four archetypes: Hero (benevolent, strong); Victim (benevolent, weak); Villain (malevolent, strong); and Fool (malevolent, weak).

\section{Methods}
\label{sec:method}

\subsection{Data collection}

To examine Ukraine's use of memetic engagement during the first 8 months of the 2022 conflict, we begin with a large collection of tweets matching war-related keywords (see \Cref{sup:keywords}) spanning from 27 February 2022 to 12 October 2022. 
As outlined in \Cref{sec:intro}, we are particularly interested in the analysis of official, government-run accounts, to study their unprecedented role in a large-scale conflict.
Thus, we identify the most popular (in terms of number of followers) non-personal accounts ran by the Ukrainian government:
\begin{enumerate}
    \item[(1)] \textbf{\ukr}: self-described as ``the official Twitter account of Ukraine'', this verified account is labeled as an ``Ukraine government organization'' by Twitter. It posts war-related and political updates about Ukraine.
    \item[(2)] \textbf{\dou}: or ``Defense of Ukraine'', is the ``Official page of the Ministry of Defense of Ukraine''. Like \ukr, this verified account is described by Twitter as a ``Ukraine government organization''. Its posts concern battlefield updates and news.
\end{enumerate}
These two accounts have often employed memes to support their message.
Our primary research goal is to investigate this activity.
Besides these two official accounts, we identify another non-personal account with an important role in terms of memetic warfare:
\begin{enumerate}
    \item[(3)] \textbf{\umf}: self-described as the ``Source of the best Ukrainian memes'', this unverified account states as its goal ''to convey the truth about what is happening now in Ukraine with the help of memes''.\footnote{https://www.patreon.com/uamemesforces}
\end{enumerate}
This meme page was created in February 2022, when Russia began the invasion of Ukraine.
A preliminary analysis of our data confirms that, by the end of our collection period, this account was the most followed Twitter page entirely dedicated to creating and spreading pro-Ukrainian memes about the conflict.
Scholars have highlighted how \umf ``is followed by academics, journalists from leading publications, diplomats from a host of countries, communications advisers to world leaders and social media managers at multilateral institutions'', and thus how its memes have the potential to ``reach global elites''.\footnote{\url{https://digdipblog.com/2022/08/08/the-power-of-memes-analyzing-war-time-messaging/}}
For these reasons, we choose to focus our in-depth analysis and labeling efforts on these three accounts, which represent the most prominent examples of memetic warfare in the Ukrainian war.

We retrieve all of the posts by these accounts via Twitter API. %
\rev{We further use the Search API to retrieve the number of retweets of each post at a later date (January 2023), so that this number has stabilized.}
As explained in \Cref{sec:relwork}, memes are defined not by any intrinsic characteristic, but by their ability to be reshared.
As such, we treat each post with visual media content as a potential meme, irrespective of the format.
\Cref{sec:annotation} illustrates how we incorporate the features that previous research has identified as typically associated with viral memes into our analysis.
To ensure preserving content posted by these accounts, we also download the images associated with the posts and the users who retweeted or liked the post.
We associate each user with a country in the GeoNames database\footnote{\url{https://www.geonames.org}} via their location field (a free-text field optionally filled by the user).
Considering only tweets with media, out of \num{586042} users, \num{44}\% are successfully mapped to a country.
For geographic analysis, we retain only countries with at least \num{1000} geolocated retweeting users.

\rev{Considering all the tweets obtained through the Search API}, \ukr consistently garners the highest level of engagement despite having only 54 tweets.
This account's tweets achieved an average of \num{9670} retweets per post, significantly exceeding \dou (\num{750} retweets per post) and \umf (\num{531} retweets per post).
\dou is the account with the lowest number of retweets, however, its audience is more dedicated and engaged.
In fact, on average a user retweets \dou approximately \num{3.60} times, compared to \num{2.38} for \umf and \num{1.18} for \ukr.
Among the three Twitter accounts analyzed, \ukr exhibits the broadest geographical reach, with a significant portion of retweets originating from outside of Ukraine.
Specifically, $27\%$ of retweets are from the United States, $0.9\%$ from the United Kingdom, and $0.8\%$ from Brazil, while only $0.013\%$ from Ukraine.
\umf garners the highest engagement from users in Japan ($19\%$) and the United States ($17\%$), while Ukrainian users contribute only $0.05\%$ of retweets.
In contrast, $14\%$ of retweets for \dou originate from Ukraine, with the remaining retweets primarily from the U.S. ($14\%$) and Japan ($13\%$).

To explore the relationship between Ukrainian memes and each country's actions in the conflict, we use the Ukraine Support Tracker data from the Kiel Institute for the World Economy,\footnote{Downloaded on 20 June 2023.} which ``quantifies military, financial, and humanitarian aid transferred by governments to Ukraine since the end of diplomatic relations between Russia and Ukraine on January 24, 2022''.\footnote{\url{https://www.ifw-kiel.de/topics/war-against-ukraine/ukraine-support-tracker}}
The dataset provides a quantitative measure of a country's aid to Ukraine.
Finally, to examine the attitude of each country's population towards Ukraine, we use the Eurobarometer STD98 survey, conducted in the winter of 2022-2023 in 30 European countries~\cite{eu2023eurobarometer}. %
Specifically, we use use QE2, answer 6: ``To what extent do you agree or disagree with: providing financial support to Ukraine''.

\newcommand{\douFractionOfAnnotatedWrtMedia}{\ensuremath{31.6\%}\xspace}
\newcommand{\umfFractionOfAnnotatedWrtMedia}{\ensuremath{39.4\%}\xspace}
\newcommand{\ukrFractionOfAnnotatedWrtMedia}{\ensuremath{94.4\%}\xspace}

\subsection{Annotation}
\label{sec:annotation}

To understand the visual and narrative attributes of the captured data, all authors perform a manual annotation of a selection of the tweets (for a total of \num{1063} tweets, see \Cref{tab:data_stats_annotated_tweet}).
We consider all tweets with media as candidates for the annotation.
We remove self-replies (an account replying to its own post) which have no retweets of their own.
For the two accounts with a large number of tweets we consider a random sample (\douFractionOfAnnotatedWrtMedia for \dou and \umfFractionOfAnnotatedWrtMedia for \umf):
this way, we balance the labeling effort with the number of annotated tweets.
Instead, given the small number of tweets with media by \ukr, we annotate all the available tweets that pass the filters (\ukrFractionOfAnnotatedWrtMedia of \ukr tweets with media).
Summary statistics for the dataset of annotated tweets (after filtering and sampling) are shown in \Cref{tab:data_stats_annotated_tweet}.

Guided by previous literature, we compose a codebook to capture visual and narrative aspects of the content. 
First, we adopt the significant features of the content characterization framework proposed by \citet{ling2021dissecting}, which is used to model the virality of political memes. 
These include:
\begin{enumerate}
    \item Number of panels: multiple or single;
    \item Type of image: photo, screenshot, or illustration;
    \item Scale: close-up, medium shot, long shot, other;
    \item Type of subject: object, character, scene, creature, text, other;
    \item Main attribute of the subject (if character): facial expression, posture, poster
    \item Character emotion (if character): positive, negative, or neutral;
    \item Contains words: whether the image itself (not tweet caption) contains words.
\end{enumerate}

\begin{table}[t]
\caption{Statistics for the dataset of annotated tweets. The averages are computed only on the subset of annotated tweets. The reported number of followers is the last available one.}
\footnotesize
\centering
\sisetup{table-number-alignment=right}
\label{tab:data_stats_annotated_tweet}
\begin{tabular}{lrrrr}
\toprule
         user & tweets & avg. likes & avg. RT & num. followers \\
\midrule
   \ukr & \num{51} & \num{33065.50} & \num{6245.68} & \num{2274857} \\
   \dou & \num{555} & \num{5316.44} & \num{852.14}  & \num{1557170} \\
   \umf & \num{457} & \num{6750.05} & \num{637.33}  & \num{321297} \\
\bottomrule
\end{tabular}
\end{table}

Second, we employ the narrative framework by \citet{karpman1968fairy} later extended by \citet{de2021internet}. 
It defines two key dimensions to characterize a narrative: moral quality (benevolent or malevolent) and power (strong or weak), thus resulting in four character archetypes: heroes (benevolent, strong), victims (benevolent, weak), villains (malevolent, strong), and fools (malevolent, weak).
We also label the actors mentioned in these narratives (as an open category, guided by an initial set of main actors in the conflict, such as Zelensky and Putin). 
We further contextualize the narratives by their emotional appeal~\cite{chagas2019political}, in particular humor, pride, fear, outrage, and compassion (the label was originally open, but the labelers coalesced on this set).
Finally, we record whether there is a specific intent in terms of a call to action or information sharing.
Note that, as much as possible, we label the narratives as intended from the point of view of the posting account (although some biases may creep in, as we discuss in \Cref{sec:discussion}). 
\Cref{sec:codebook} reports the full codebook.

Using this literature-driven codebook, all authors of this manuscript perform a deductive coding of the selected posts (for more on such process, see \citet{linneberg2019coding}). 
We rely on Google Translate to translate non-English language content.
All coders first annotated a selection of 50 posts, discussed insights and disagreements, and came up with standard labels for emotional appeal, intent, and actor (though the labels remained open).
\rvf{During this process, the coders referred to the coding schemes from previous literature~\cite{ling2021dissecting} to adapt the concepts to the task at hand. In particular, identifying the presence or absence of an actor proved to be a challenging task.
For instance, \dou posts often represented people who may be civilians, combatants, or officials, whose roles may be difficult to distinguish (indeed, this label had the lowest agreement).
Concerning the narrative, often disagreement arises on whether one or several narratives are present in the content.
For instance, when both parties are represented (or even implied), an image can be labeled as focused on the perceived victim, the villain, or both.
}
The remaining posts were then annotated by one author each, with uncertain cases discussed collectively.
\Cref{sec:agreement} reports the inter-annotator agreement on a selection of 48 posts (sampled from each account).
On average, the agreement \rev{(as measured by Krippendorff's alpha)} is \num{0.750} for \ukr, \num{0.604} for \umf, and \num{0.667} for \dou (with \rev{an average value} of around \num{0.65} for all accounts and tasks).
These values are high considering the subjective nature of the tasks and that some features were open and could have multiple labels. 
The annotated dataset is available at \datasetUrl. %

\subsection{Popularity model}
Using the data collected and annotated as per the previous sections, we construct models to find whether there are trends in ($i$) what kind of content becomes more popular, and ($ii$) whom it reaches.
Specifically, we use linear regression to model the log-transformed number of retweets via the content variables outlined above as independent variables.
As the three accounts have posted a vastly different number of tweets, we weight the data points (tweets) inverse-proportionally to their representation in the dataset, so that each account weighs equally on the outcomes.
We build several versions of the model to determine the benefit of each class of variables: visual~\cite{ling2021dissecting}, actor, emotion, intent, and narrative. 
A baseline model simply uses the identity of the author (i.e., group fixed effects), the (log-transformed) number of followers of that account \rev{at the time of posting the meme} \rvf{(available thanks to the Streaming API v1.1)\footnote{\url{https://developer.x.com/en/docs/x-api/v1}}} as a proxy for the potential audience, and the number of days from the invasion to account for changing interest in the topic over time.
\rev{We then select the best model according to the Bayesian Information Criterion (BIC), a model selection metric that captures both unexplained variation in the dependent variable and the number of explanatory variables.
The coefficients of this model show the associations between these variables and the virality of the content.}

Next, we group the retweets by the country of the user (those that can be geolocated) to understand the preferences of users who spread these messages. 
We enrich our characterization of the relationships between each country and Ukraine by considering the amount of financial assistance the country provides to Ukraine (normalized by the country's GDP) or by their opinion on this topic (via the Eurobarometer survey).
As mentioned above, we consider these measures as different proxies of support for Ukraine's military efforts.
First, we examine the retweet rate (normalized by population).
Then, we compare the narrative preferences to the amount of support for Ukraine: are there narratives associated with increased support for Ukraine?
We operationalize this preference as the log odds of users from a country retweeting content with a specific narrative.
Finally, we examine the geographical distribution of the retweet rate and narrative preferences across Europe by plotting theme on choropleth maps.

\section{Results}

We now analyze the annotated dataset to answer our research questions.
\Cref{sec:first-statistics} outlines which narratives, emotions, and actors are more prevalent in each of the three accounts.
\Cref{sec:popularity} examines which factors are more important for the virality of this content.
Finally, \Cref{sec:audience} shows how the audience breaks down across countries. %

\begin{figure}[tb]
	\centering
	\begin{subcaptionblock}[T]{0.32\textwidth}
		\includegraphics[width=\textwidth]{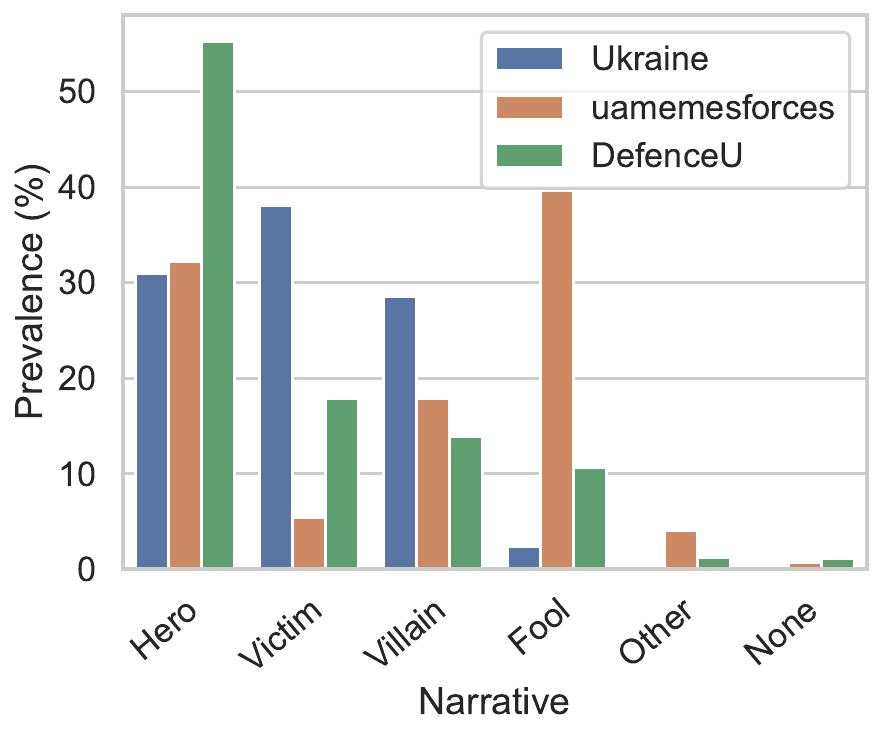}
	\end{subcaptionblock}
	\begin{subcaptionblock}[T]{0.32\textwidth}
		\includegraphics[width=\textwidth]{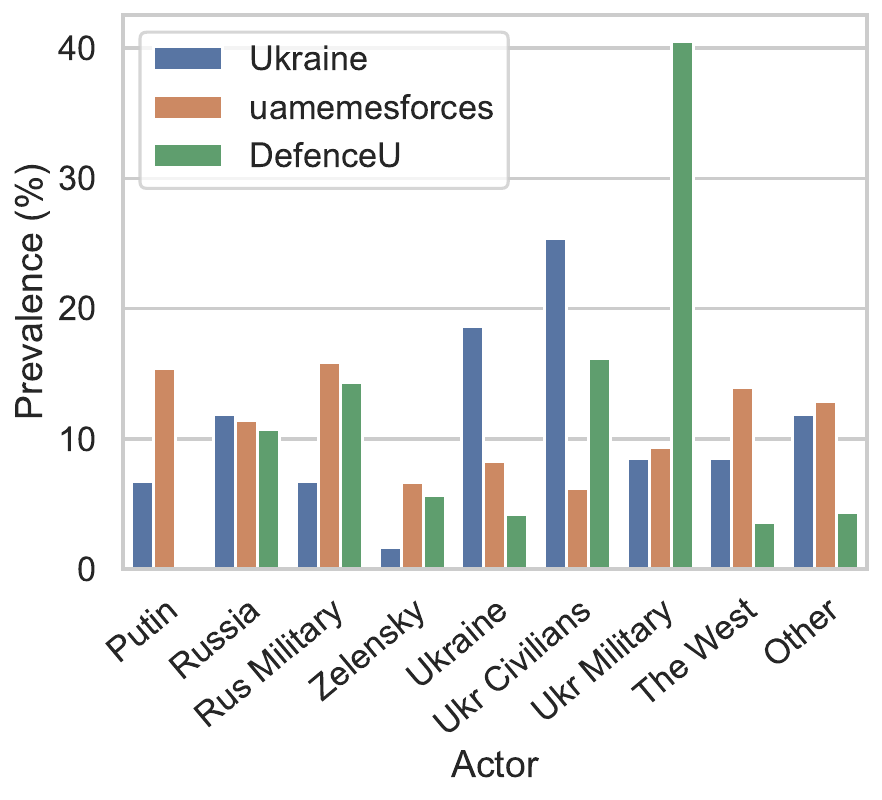}
	\end{subcaptionblock}
	\begin{subcaptionblock}[T]{0.32\textwidth}
		\includegraphics[width=\textwidth]{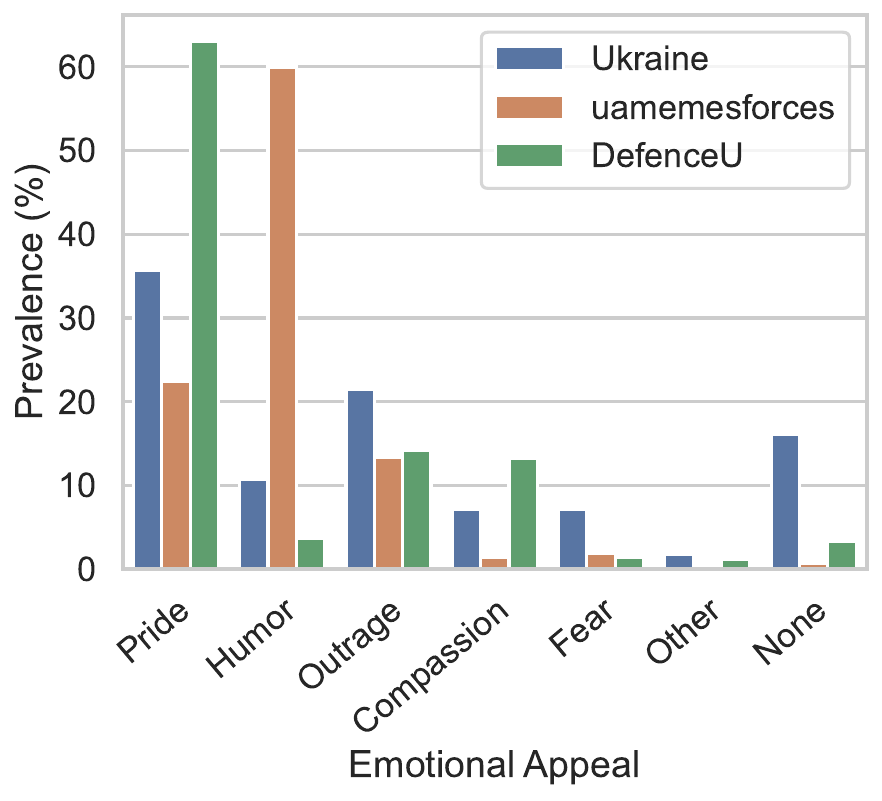}
	\end{subcaptionblock}
	\caption{Distributions of narratives, actors, and emotional appeals in the messages of the three Ukrainian accounts. The accounts show marked differences in their communication strategies. \ukr uses a balanced mix, with an emphasis on civilians, but rarely makes fun of their target. Conversely, \umf favors the fool narrative and uses humor generously. Finally, \dou favors the hero narrative, mostly evoking a sense of pride for Ukraine's military.}
	\label{fig:narrative-statistics}
	\vspace{-.5\baselineskip}
\end{figure}

\subsection{Content analysis}
\label{sec:first-statistics}

\Cref{fig:narrative-statistics} outlines which elements are more prevalent in each of the considered accounts.
\ukr uses a mix of narratives but completely avoids the fool one: portraying the enemy as incompetent is counter-productive when asking for help\rev{, something already known since World War 2~\citep{herz1949psychological}}.
Unsurprisingly, \dou focuses on the heroic narrative by portraying the Ukrainian military as powerful and morally good.
Instead, the independent account \umf focuses on the fool narrative, in which the other side is portrayed as weak and incompetent.
At the same time, the account employs a large amount of hero narrative, partially due to numerous memes of Ukrainian farmers resisting Russian soldiers.

The actors mentioned by each account also differ.
\ukr focuses the most on Ukrainian civilians and Ukraine as a nation, possibly working towards the goal of nation-building, or to keep the global public opinion focused on the main victims of the conflict.
\dou puts the spotlight on the Ukrainian military but also dedicates some attention to Ukrainian civilians.
Interestingly, Putin is mentioned by \umf much more than \dou, which almost never mentions him.
Similarly, actors associated with the West are mentioned more by \umf, while \dou tends to ignore them.
This result is in line with the more domestic focus of \dou we previously highlighted.
The three accounts are also distinct in their emotional appeal: \ukr communicates largely outrage and pride, and \dou pride alone, while \umf focuses on humor. 
Concerning possible intents (not plotted), \ukr posts by far the most calls to action (31\%), with many calls to support the war effort, while \dou often posts purely informational material (e.g., war statistics bulletins).

\begin{figure}
	\centering
	\begin{subcaptionblock}[T]{0.32\textwidth}
		\includegraphics[width=\textwidth]{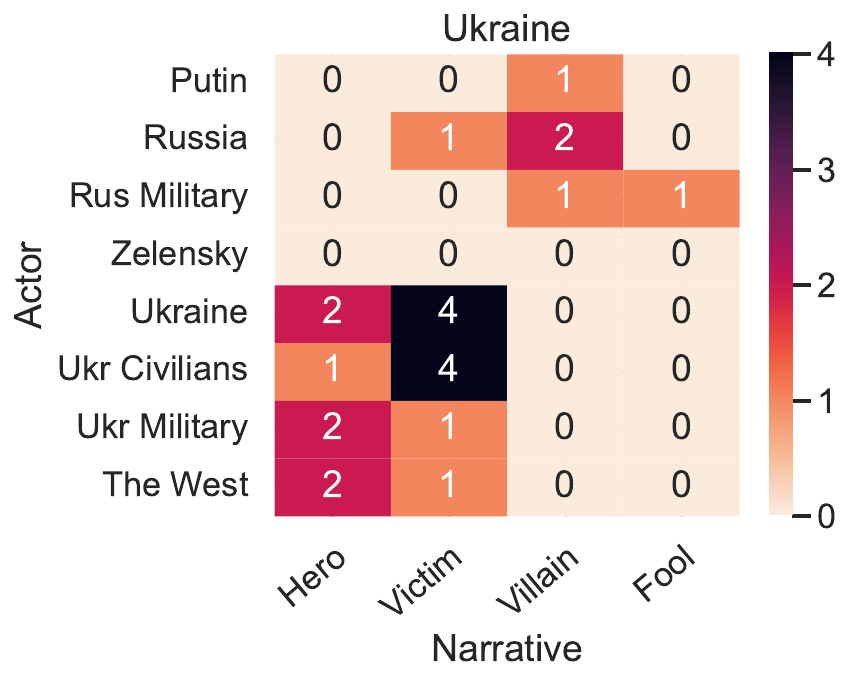}
	\end{subcaptionblock}
	\begin{subcaptionblock}[T]{0.32\textwidth}
		\includegraphics[width=\textwidth]{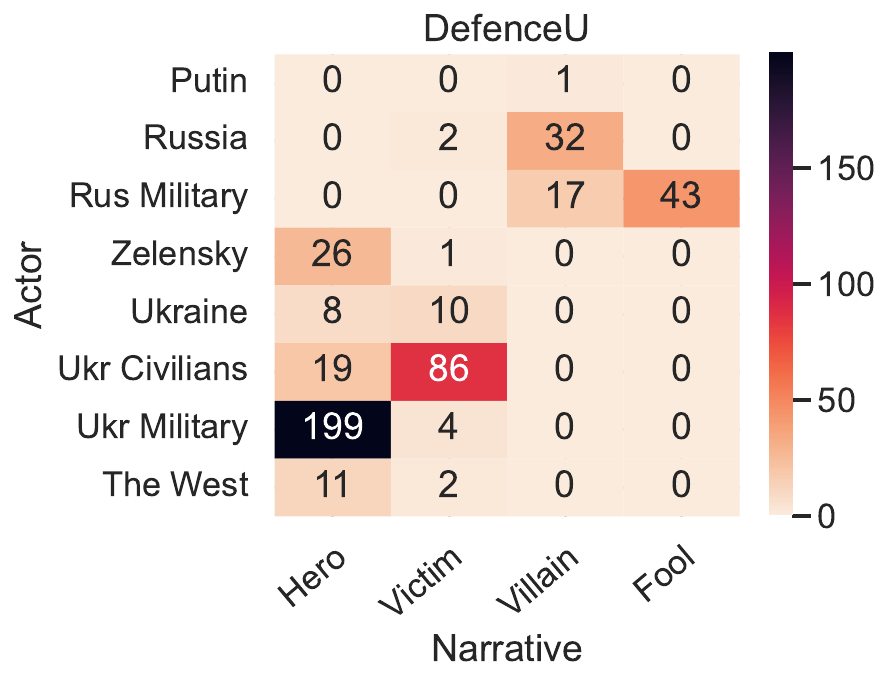}
	\end{subcaptionblock}
	\begin{subcaptionblock}[T]{0.32\textwidth}
		\includegraphics[width=\textwidth]{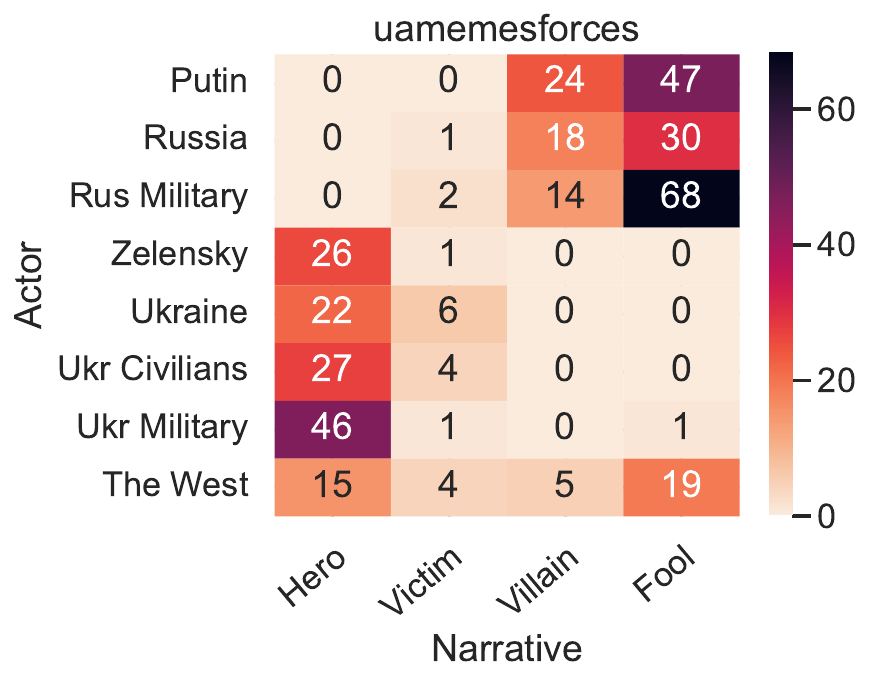}
	\end{subcaptionblock}
	\caption{Distributions of co-mentions of specific actors along with a given narrative for the three Ukrainian accounts. The block structure of the 2d distribution shows that malevolent narratives are used when talking about Russian actors, while benevolent ones when talking about Ukrainian actors. \ukr puts an emphasis on the victim narrative and \dou has just two clear narratives that emerge (Ukrainian military as heroes and civilians as victims). \umf instead focuses most of its messaging on making fun of and Russian military and Putin.}
	\label{fig:actor_narrative}
	\vspace{-.5\baselineskip}
\end{figure}

\begin{figure}
	\centering
	\begin{subcaptionblock}[T]{0.32\textwidth}
		\includegraphics[width=\textwidth]{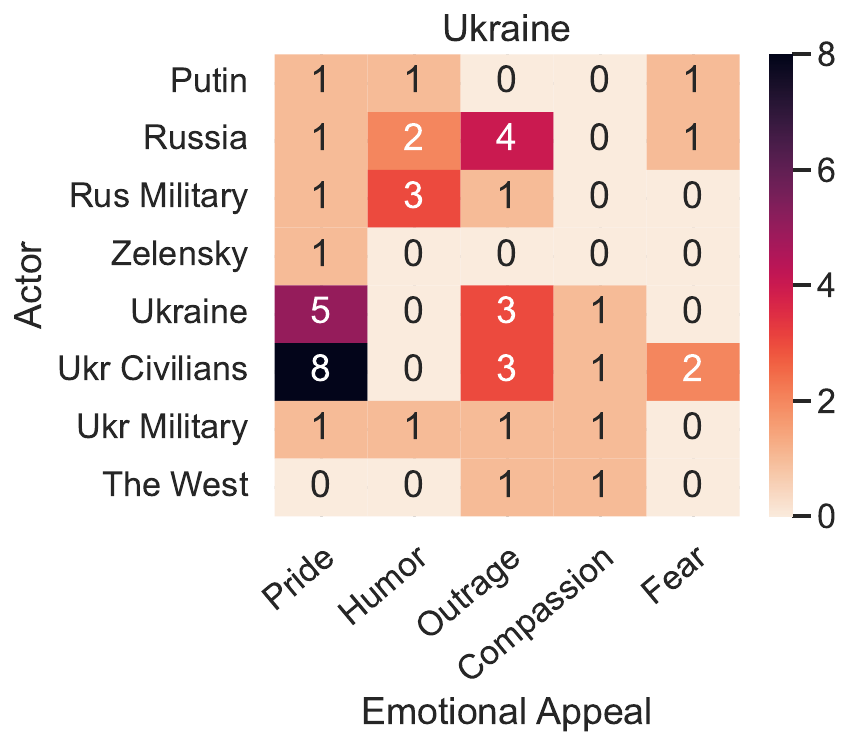}
	\end{subcaptionblock}
	\begin{subcaptionblock}[T]{0.32\textwidth}
		\includegraphics[width=\textwidth]{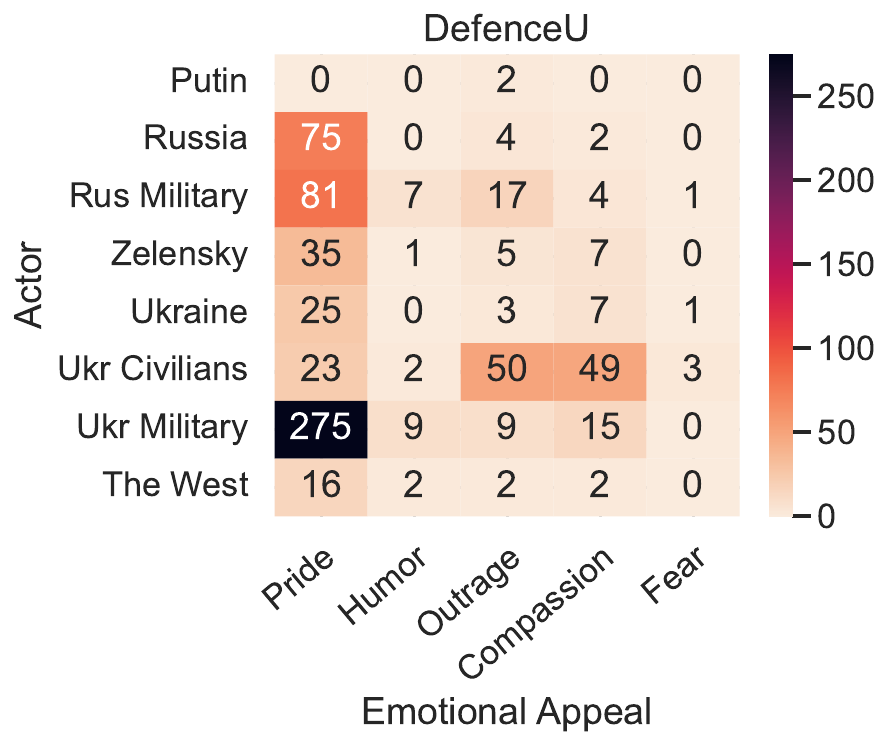}
	\end{subcaptionblock}
	\begin{subcaptionblock}[T]{0.32\textwidth}
		\includegraphics[width=\textwidth]{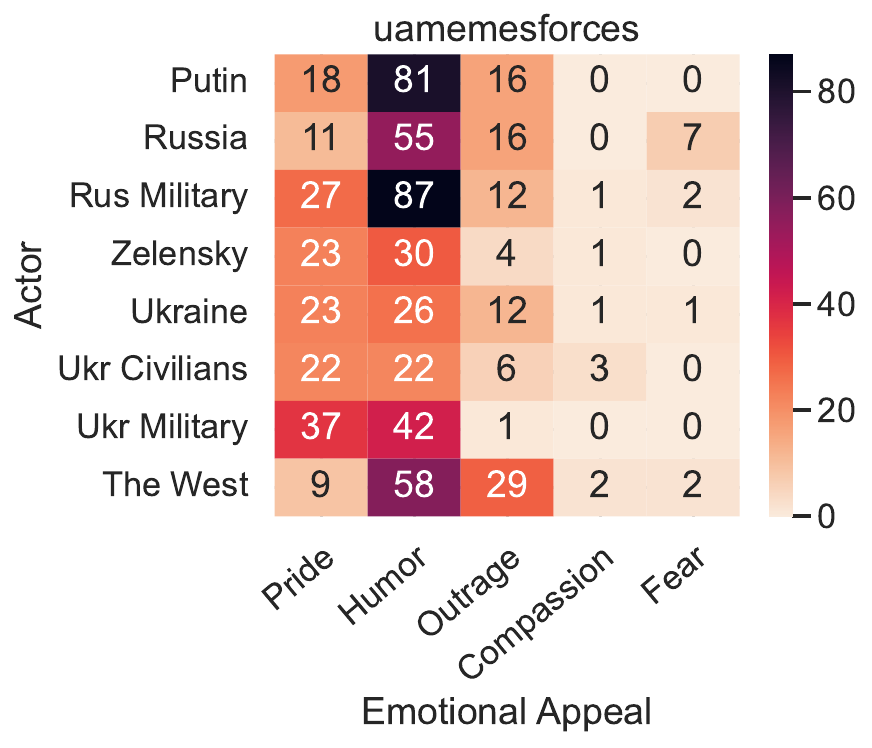}
	\end{subcaptionblock}
	\caption{Distribution of co-mentions of specific actors along with a given emotional appeal. While clear preferences for distinct rhetorical styles are present across the accounts, they do not show consistent associations with specific actors.}
	\label{fig:actor_emotions}
	\vspace{-.5\baselineskip}
\end{figure}

Next, we examine the role of the main actors found in the narratives.
\Cref{fig:actor_narrative} presents heat maps of the intersection of actors and narratives for each account.
Across all three accounts, the dichotomy between the two sides of the conflict is stark: the Russian side is portrayed as either a fool or a villain (negative moral quality), while the Ukrainian side as either a hero or a victim (positive moral quality); the West spans both portrayals. 
The emphasis by \ukr is on Ukrainian civilians and Ukraine as a whole, portrayed in a victim role, whereas less emphasis is put on the actors from the Russian side.
The narrative of \dou consists of several well-defined combinations: Ukrainian military as heroes, Russian military as fools, and Ukrainian civilians as victims.
It is noteworthy that the Russian military is often portrayed as powerless (fool), while Russia as a country is always portrayed as a villain (powerful), and Putin himself is rarely referred to personally (contrary to the practice of \umf).
Unlike for \ukr, \umf content rarely portrays Ukraine as a victim, even when civilians are involved, and instead attributes them heroic characteristics: recurring examples of this combination are references to a story of Ukrainian farmers who towed away abandoned Russian tanks.
The most common combination of actor/narrative is portraying the Russian military as a fool (negative moral quality and powerless), and a similar treatment is reserved for Putin.
This powerless portrayal is much more common than the powerful one (in the role of a villain), and is in line with the satyrical role of the account.
Actors from the West are mostly portrayed as fools or heroes---mostly depending on whether the particular actor is perceived as supporting Ukraine---but in a few cases, they are even portrayed as villains, when they are perceived to be complicit with the enemy. %
Russian civilians are only briefly mentioned by \umf (omitted from the plots, 20 tweets in total), often as ``conniving'' with their government. %
Thus, we find distinctive narrative styles in the three accounts: \ukr focuses on its victimhood, \dou propagates a heroic, powerful view of its own military while portraying its enemy in a foolish, weak perspective, and \umf emphasizes the weakness (foolishness) of the Russian side and the heroism of Ukrainians.

Finally, \Cref{fig:actor_emotions} shows the emotional appeal of the posts in each account.
Since a tweet labeled with a single emotion could refer to several actors, in our analysis we associate such emotion with all the actors present.
The emotional focus is very well defined for \dou, where pride is present in an overwhelming majority of the posts.
A combination of humor dominates the messaging style of \umf.
Interestingly, Putin is singled out as an object of ridicule only by \umf.
No clear pattern is present for \ukr.

\spara{Examples.}
Let us now focus on the narratives, given their explanatory power that we explore in \Cref{sec:popularity}.
Here we give a few examples to crystallize the imagery associated with each narrative.
These examples are taken from the top 10 most popular tweets per narrative.

\begin{figure}[tb]  %
\centering
	\begin{subcaptionblock}[t]{0.32\textwidth}
		\includegraphics[width=\textwidth]{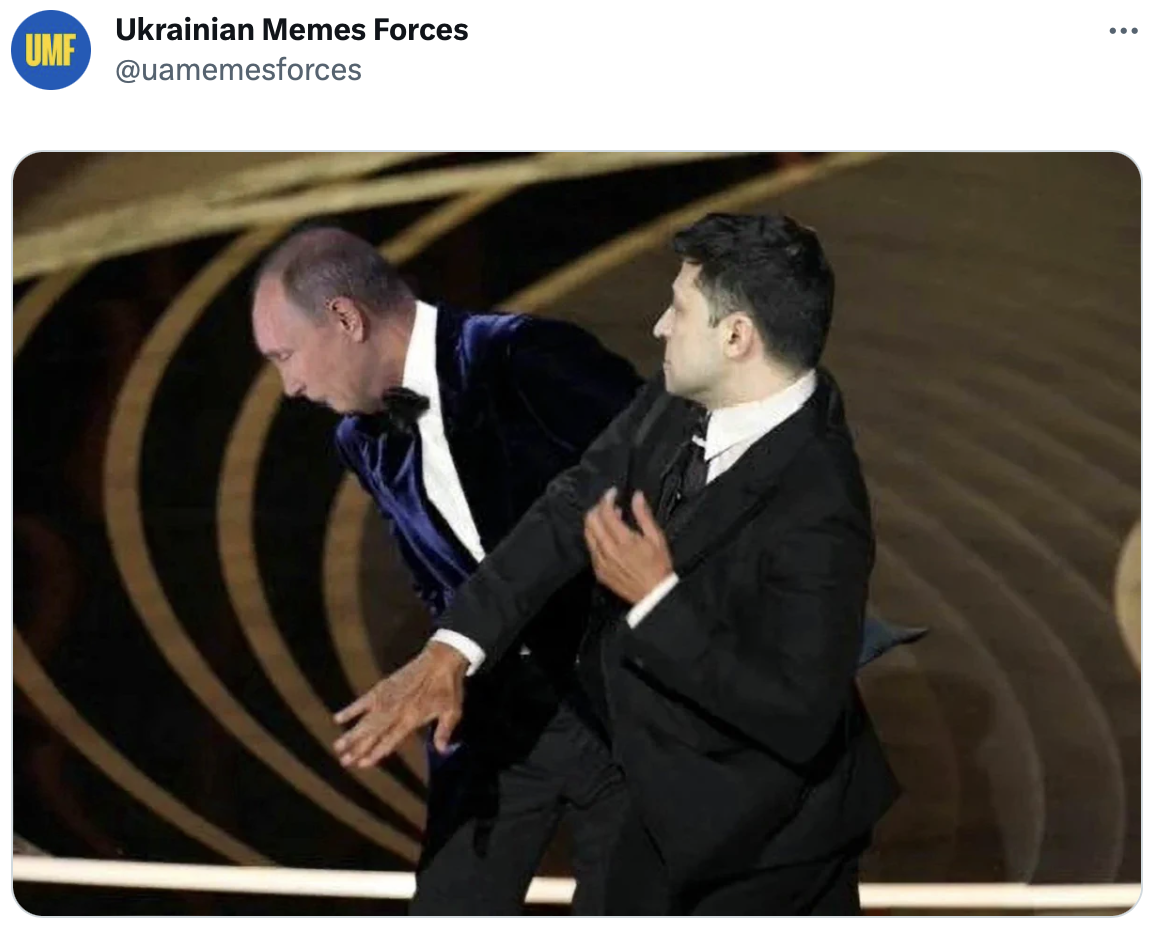}
	\end{subcaptionblock}
	\begin{subcaptionblock}[t]{0.32\textwidth}
		\includegraphics[width=\textwidth]{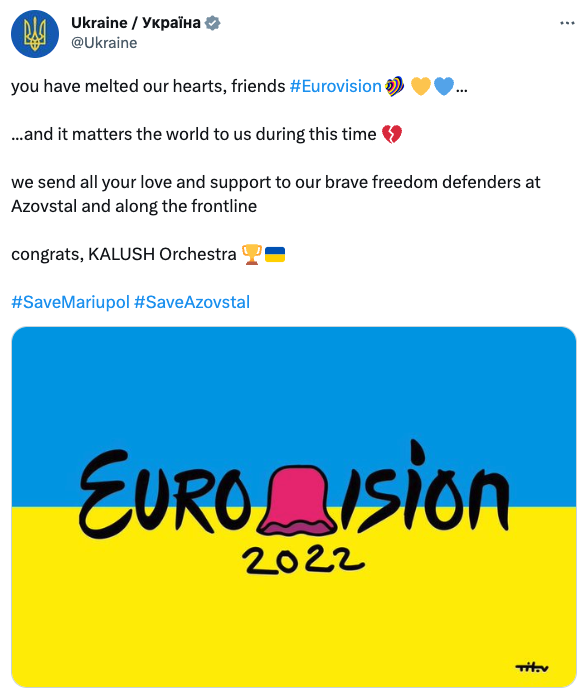}
	\end{subcaptionblock} 
	\begin{subcaptionblock}[t]{0.32\textwidth}
		\includegraphics[width=\textwidth]{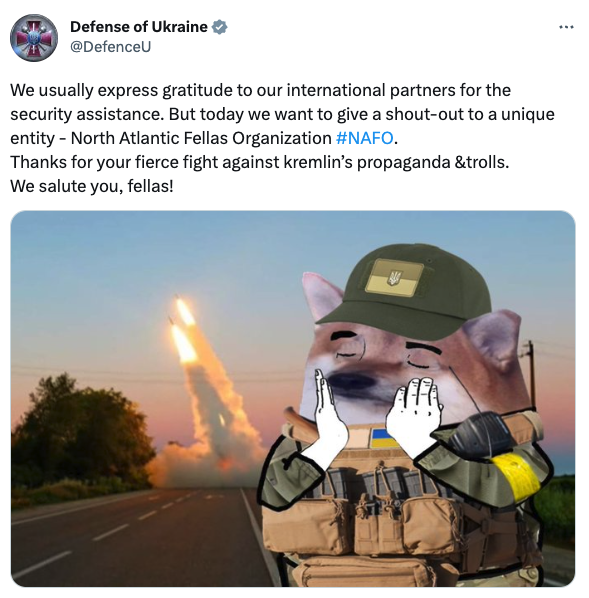}
	\end{subcaptionblock}
	\caption{Examples of `hero' narrative.}
	\label{fig:examples_hero}
	\vspace{-.5\baselineskip}
\end{figure}

\begin{figure}[tb]  %
\centering
	\begin{subcaptionblock}[t]{0.32\textwidth}
		\includegraphics[width=\textwidth]{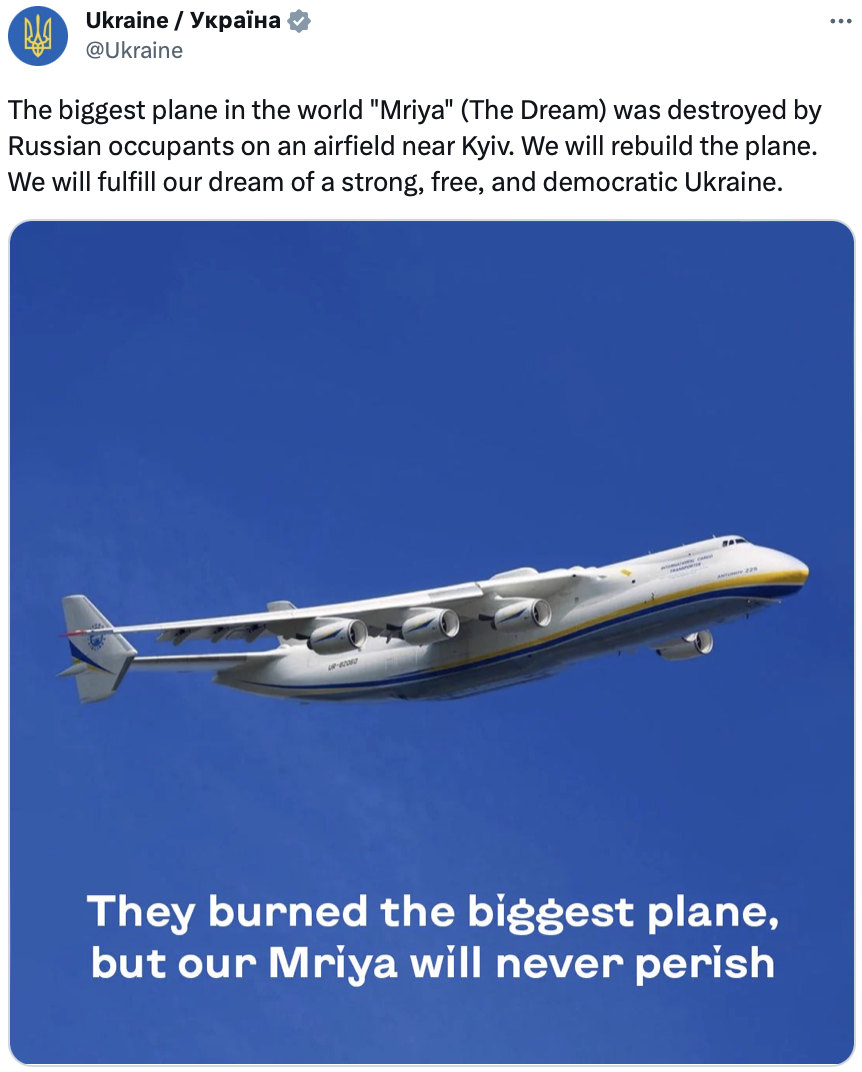}
	\end{subcaptionblock}
	\begin{subcaptionblock}[t]{0.32\textwidth}
		\includegraphics[width=\textwidth]{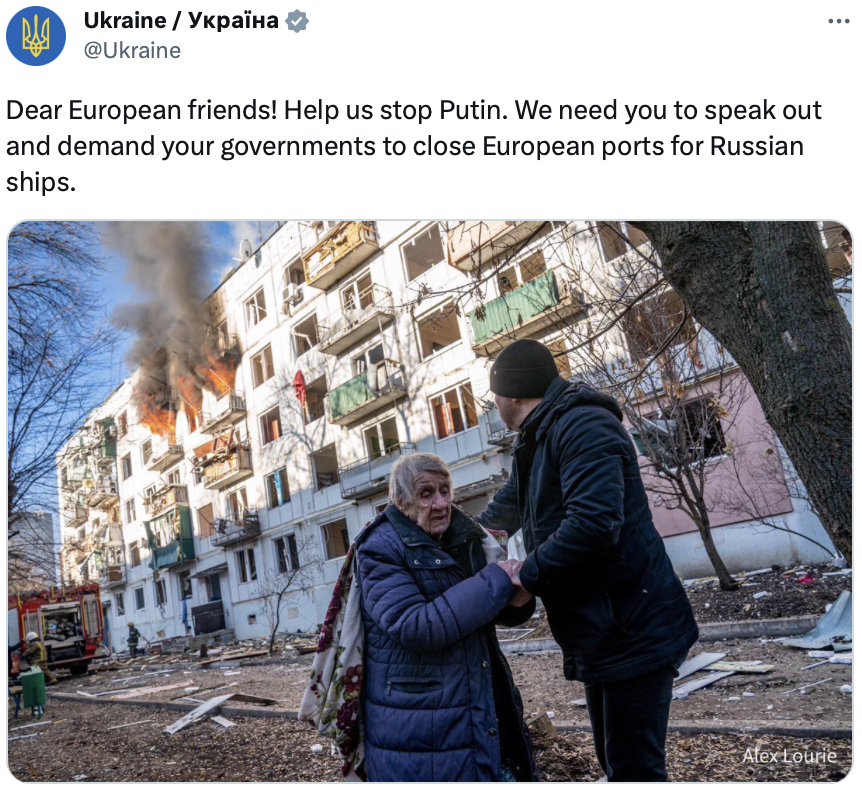}
	\end{subcaptionblock}
	\begin{subcaptionblock}[t]{0.32\textwidth}
		\includegraphics[width=\textwidth]{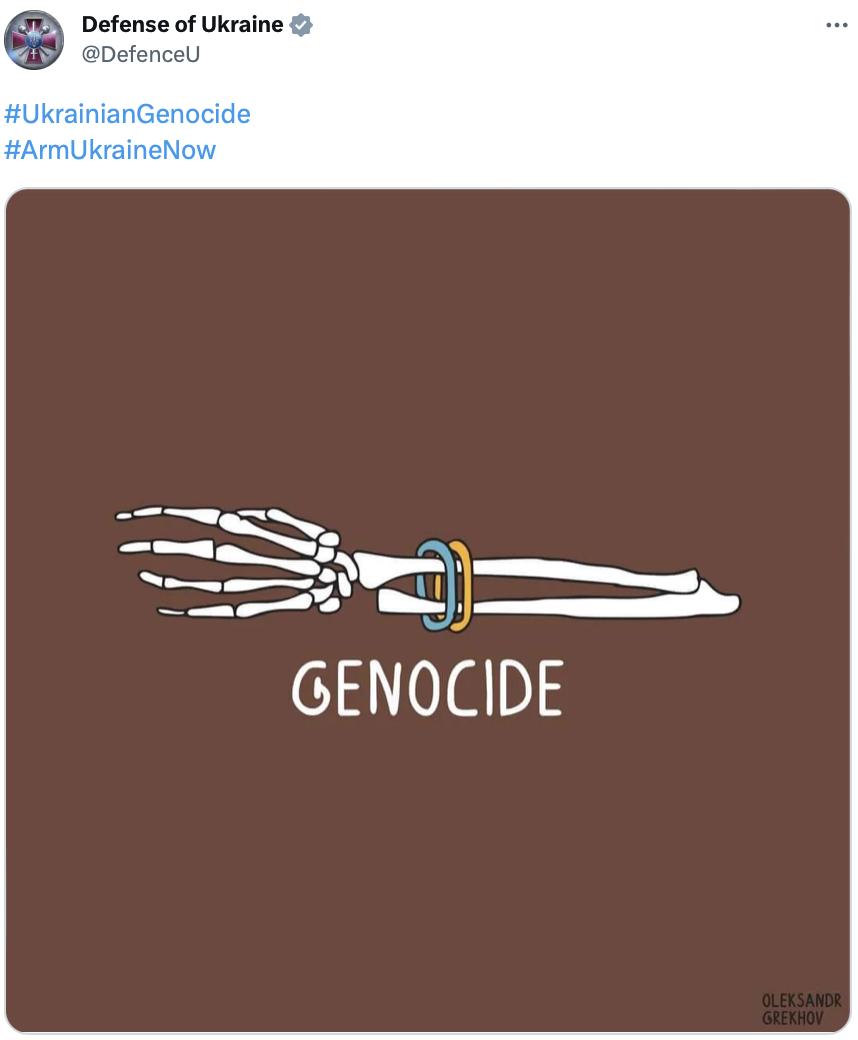}
	\end{subcaptionblock}
	\caption{Examples of `victim' narrative.}
	\label{fig:examples_victim}
	\vspace{-.5\baselineskip}
\end{figure}

\begin{figure}[tb]  %
\centering
	\begin{subcaptionblock}[t]{0.32\textwidth}
		\includegraphics[width=\textwidth]{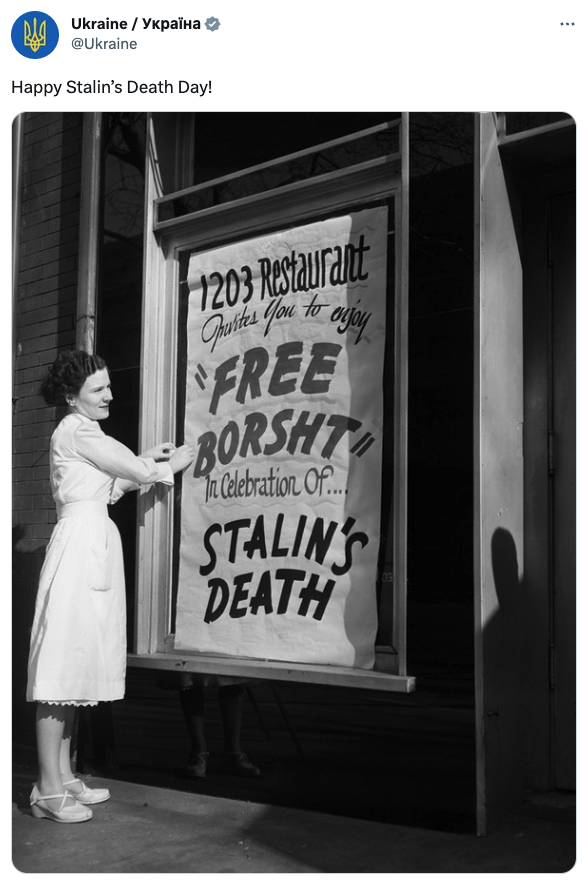}
	\end{subcaptionblock}
	\begin{subcaptionblock}[t]{0.32\textwidth}
		\includegraphics[width=\textwidth]{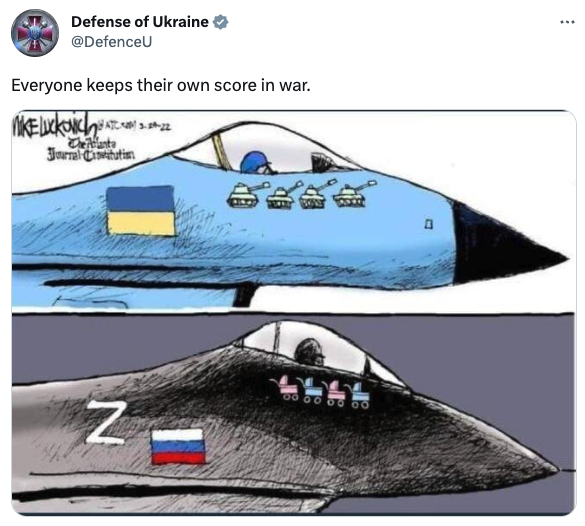}
	\end{subcaptionblock}
 	\begin{subcaptionblock}[t]{0.32\textwidth}
		\includegraphics[width=\textwidth]{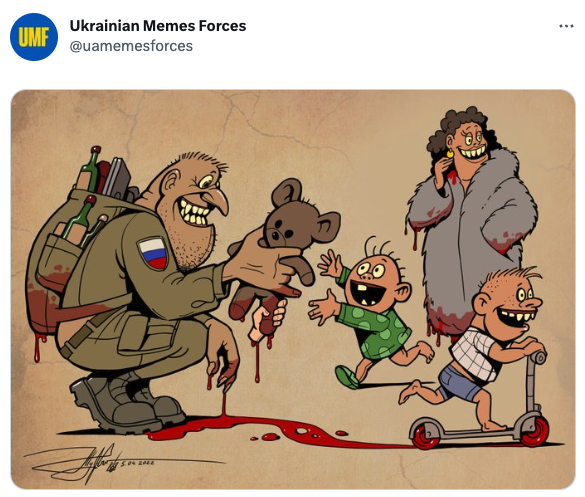}
	\end{subcaptionblock}
	\caption{Examples of `villain' narrative.}
	\label{fig:examples_villain}
	\vspace{-.5\baselineskip}
\end{figure}

\begin{figure}[tb]  %
\centering
    \begin{subcaptionblock}[t]{0.32\textwidth}
		\includegraphics[width=\textwidth]{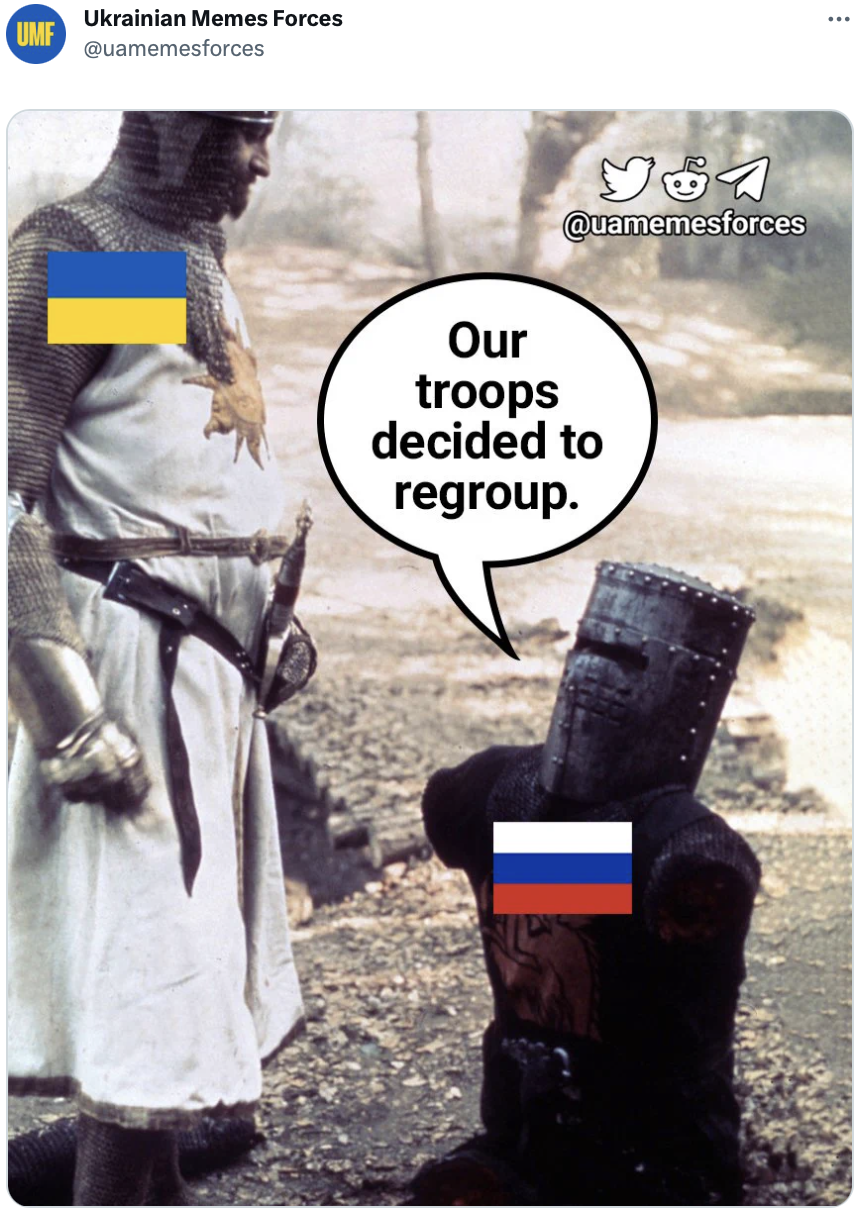}
	\end{subcaptionblock}
	\begin{subcaptionblock}[t]{0.32\textwidth}
		\includegraphics[width=\textwidth]{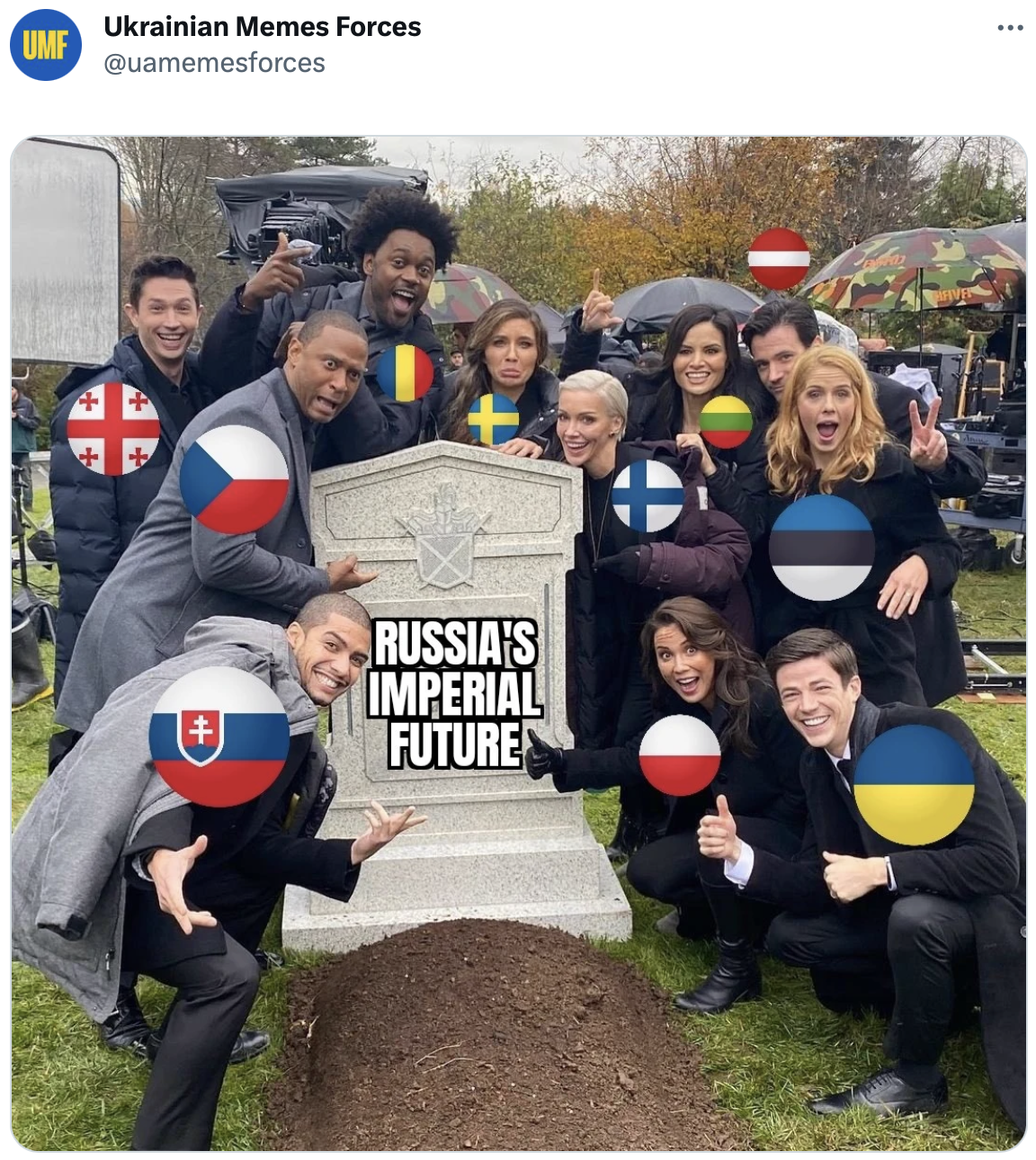}
	\end{subcaptionblock}
	\begin{subcaptionblock}[t]{0.32\textwidth}
		\includegraphics[width=\textwidth]{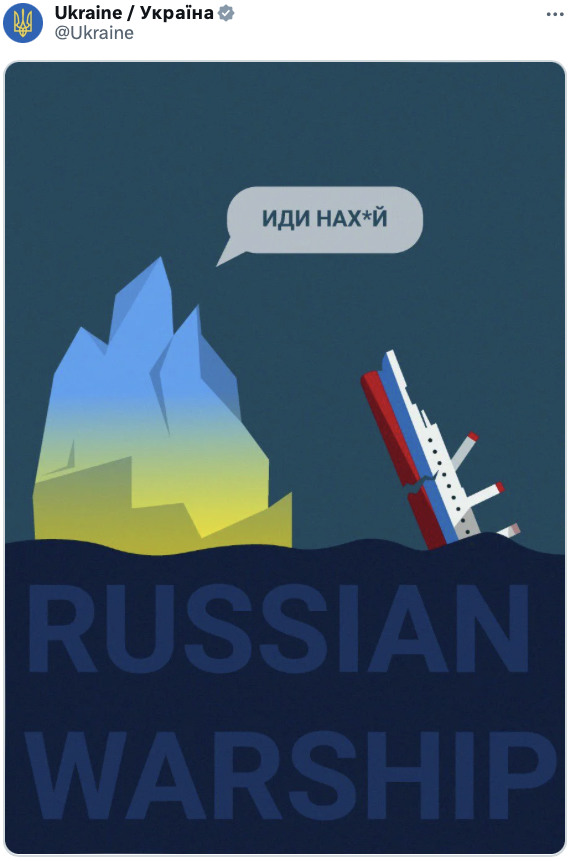}
	\end{subcaptionblock}
	\caption{Examples of `fool' narrative.}
	\label{fig:examples_fool}
	\vspace{-.5\baselineskip}
\end{figure}

\begin{figure}[tb]  %
\centering
	\begin{subcaptionblock}[t]{0.32\textwidth}
		\includegraphics[width=\textwidth]{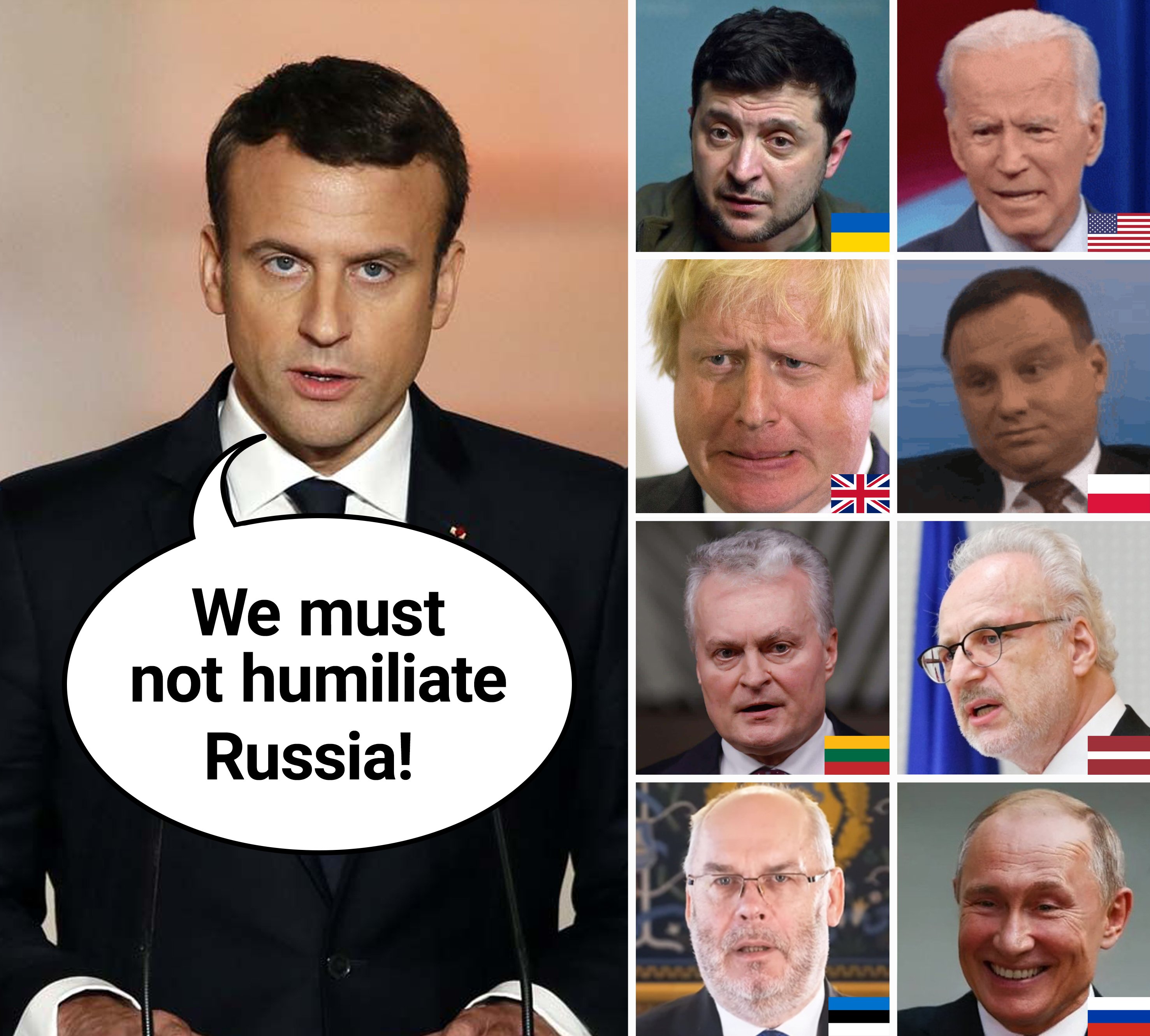}
	\end{subcaptionblock}
	\begin{subcaptionblock}[t]{0.32\textwidth}
		\includegraphics[width=\textwidth]{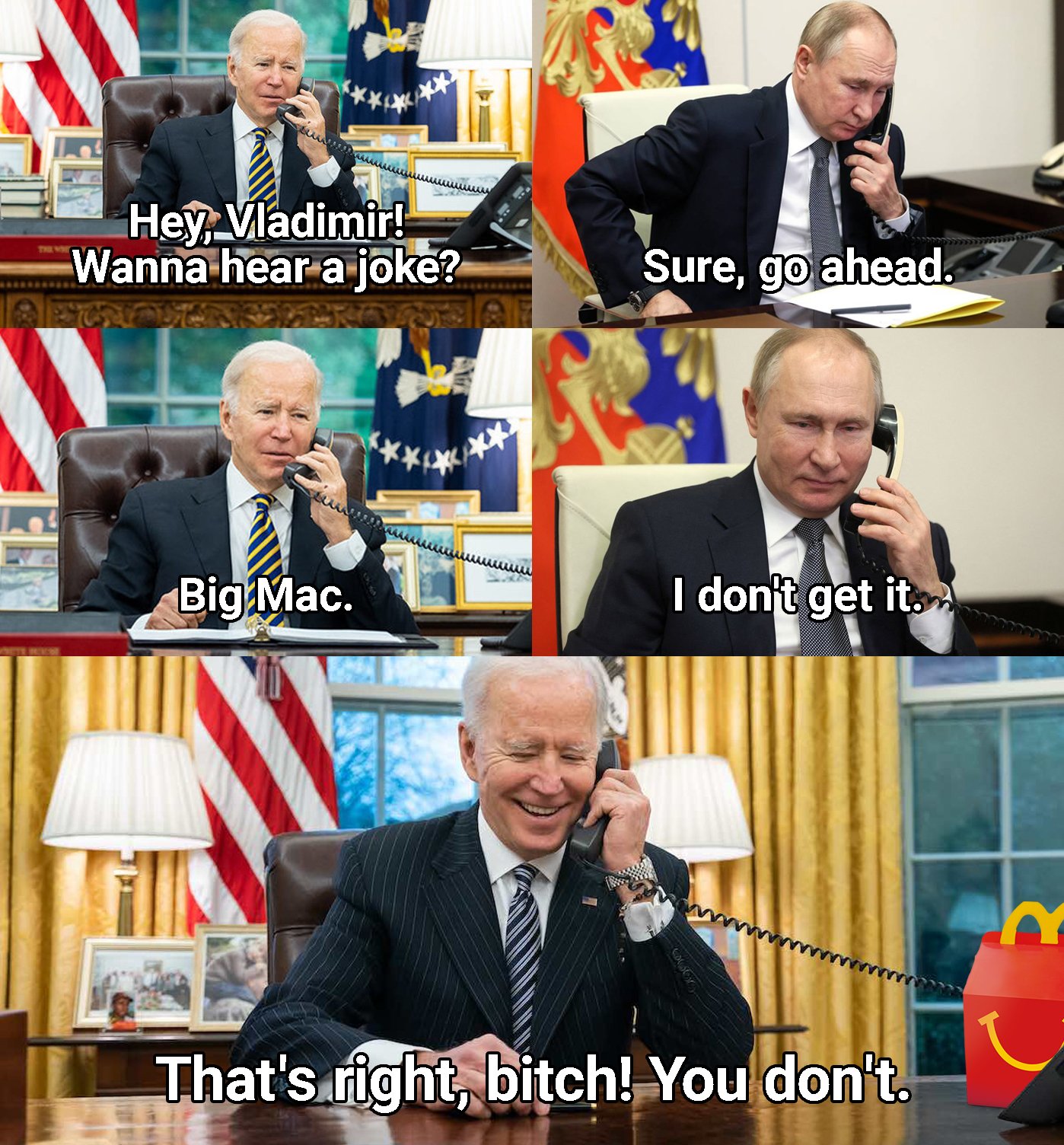}
	\end{subcaptionblock}
	\begin{subcaptionblock}[t]{0.32\textwidth}
		\includegraphics[width=\textwidth]{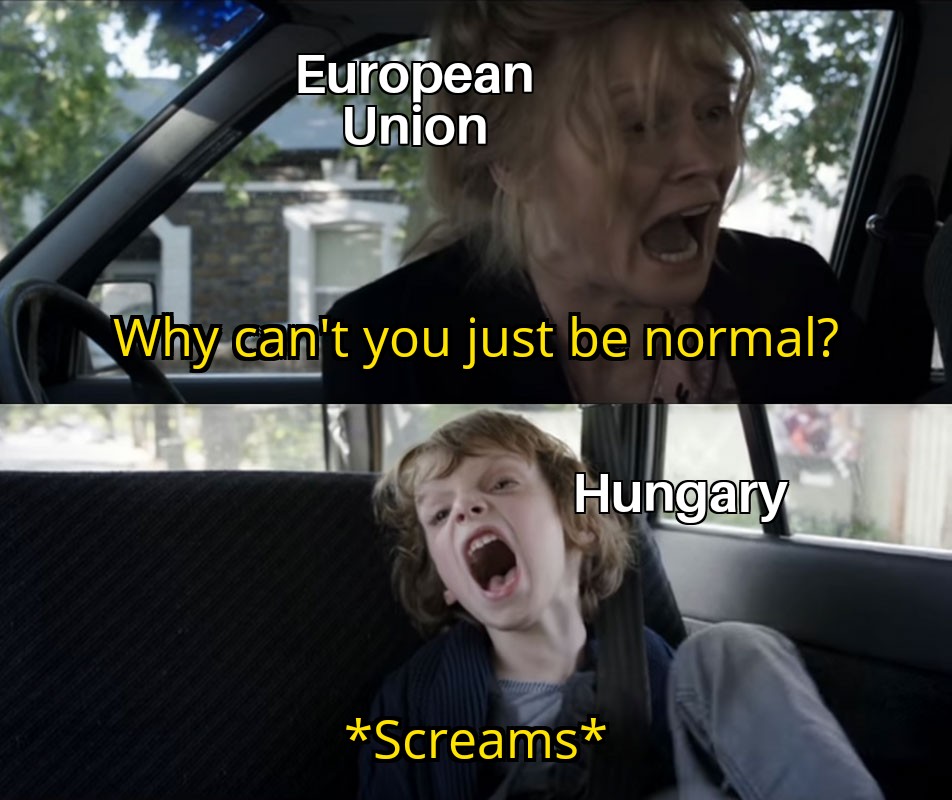}
	\end{subcaptionblock}
	\caption{Examples of `other' narrative.}
	\label{fig:examples_other}
	\vspace{-.5\baselineskip}
\end{figure}

\Cref{fig:examples_hero} illustrates examples of `hero' narrative.
This narrative typically depicts the heroism of Ukrainian military efforts, emphasizing their benevolent moral quality and their power to defeat the enemy.
In fact, it is often associated with mentions of Ukraine as an actor.
A typical example is the leftmost in \Cref{fig:examples_hero}.
This meme was posted by \umf on 28 March 2023; in the same days, Ukrainian officials were claiming victories against the Russian military forces, such as recapturing the city of Irpin. %
This meme by \umf supports the same narrative, but uses typical memetic language: the famous moment at the 2022 Oscars when Will Smith slapped Chris Rock is edited to show Zelensky fiercely slapping Putin.
This tweet obtained more than \num{120}k likes.
Other content, typically by \ukr, promotes information on ways to support Ukraine, or celebrates the valor of its civilians, such as their victory in the Eurovision Song Contest, and connects it to the ``brave freedom defenders'' on the frontline (middle example in the figure).
\dou focuses on the military: not-so-veiled military threats (e.g., a video of a military toy floating by the Kerch bridge, not shown) %
and references to an online social-media organization dedicated to countering Russian online propaganda (the ``North Atlantic Fellas Organization \#NAFO'', rightmost image).%

\Cref{fig:examples_victim} shows examples of `victim' narrative.
The most popular are from \ukr: for example, the news that the world's largest plane, the Ukrainian Antonov AN-225, had been destroyed by the Russians is used to illustrate the suffering of the war (leftmost).
Another example is shown in the middle, where \ukr uses a photo of civilians whose house was destroyed by Russian bombings in the city of Chuhuiv, near Kharkiv, to ask the global public opinion support for a specific economic sanction against Russia---the closing of seaports.
The last example is from \dou, who uses a poster-like design to promote their view that Russia's actions in Ukraine fall under the definition of genocide---something that the United Nations is currently investigating and that at the time of this writing is inconclusive.\footnote{https://www.reuters.com/world/europe/russia-has-committed-wide-range-war-crimes-ukraine-un-inquiry-finds-2023-03-16} %

\Cref{fig:examples_villain} displays examples of `villain' narrative, focusing on the immorality of the enemy and its strength.
They often juxtapose Russia with Ukraine and emphasize the civil casualties incurred by Ukraine.
Negative historical associations (e.g., with Stalin or Hitler) are often employed (leftmost example).
Extreme evilness is attributed to the Russian army: in the middle example, a cartoon shows Russian pilots gloating as they kill Ukrainian babies, a typical atrocity frame employed in propaganda~\cite{cull2003propaganda}.
Even Russian civilians are portrayed as evil: in the rightmost example, they happily play with bloodied toys taken from Ukrainian children.

\Cref{fig:examples_fool} presents examples of `fool' narrative, which still focuses on the immorality of the enemy, but depicts it as weak and powerless.
Many examples focus on Russia's failed imperial ambitions and portray its soldiers as incompetent and ineffective, as is typical of psychological warfare.

Tweets containing `other' narratives are often associated with humorous posts and mention actors other than Ukraine, bearing an ambivalent attitude.
\Cref{fig:examples_other} shows several examples, which provide a commentary on the international response to the war, the sanctions, and the role of Hungary in the conflict. 
While an overarching narrative can be inferred, it does not clearly fit within the narrative framework we employ.

\begin{table}[b]
\caption{Difference in Bayesian Information Criterion score from the baseline model ($\Delta$BIC) and Adjusted $R^2$ for models with different sets of features. More negative $\Delta$BIC indicates better models. The best model within the group is bolded. }
\label{tab:model_improvements}
\vspace{-.5\baselineskip}
\begin{tabular}{lrr}
\toprule
{Model (feature groups)} &   $\Delta$BIC &  Adj. R2 \\
\midrule
Baseline                            &   0.000 &   0.557 \\
Visual                          & -10.080 &   0.599 \\
Actor                           &  28.257 &   0.562 \\
Emotion                         & -21.336 &   0.578 \\
Intent                          &   3.480 &   0.560 \\
\textbf{Narrative}              & \textbf{-28.299} &   0.580 \\
\cmidrule(lr){1-3}
Visual+Actor                    &   7.230 &   0.608 \\
Visual+Emotion                  & -18.052 &   0.613 \\
Visual+Intent                   &  -5.742 &   0.602 \\
\textbf{Visual+Narrative}       & \textbf{-43.965} &   0.622 \\
Visual+Narrative+Emotion        & -38.423 &   0.631 \\
Visual+Narrative+Emotion+Intent & -26.376 &   0.630 \\
\bottomrule
\end{tabular}
\vspace{-.5\baselineskip}
\end{table}

\subsection{Modeling popularity}
\label{sec:popularity}

\begin{figure}[t]  %
\centering
\includegraphics[width=0.75\textwidth]{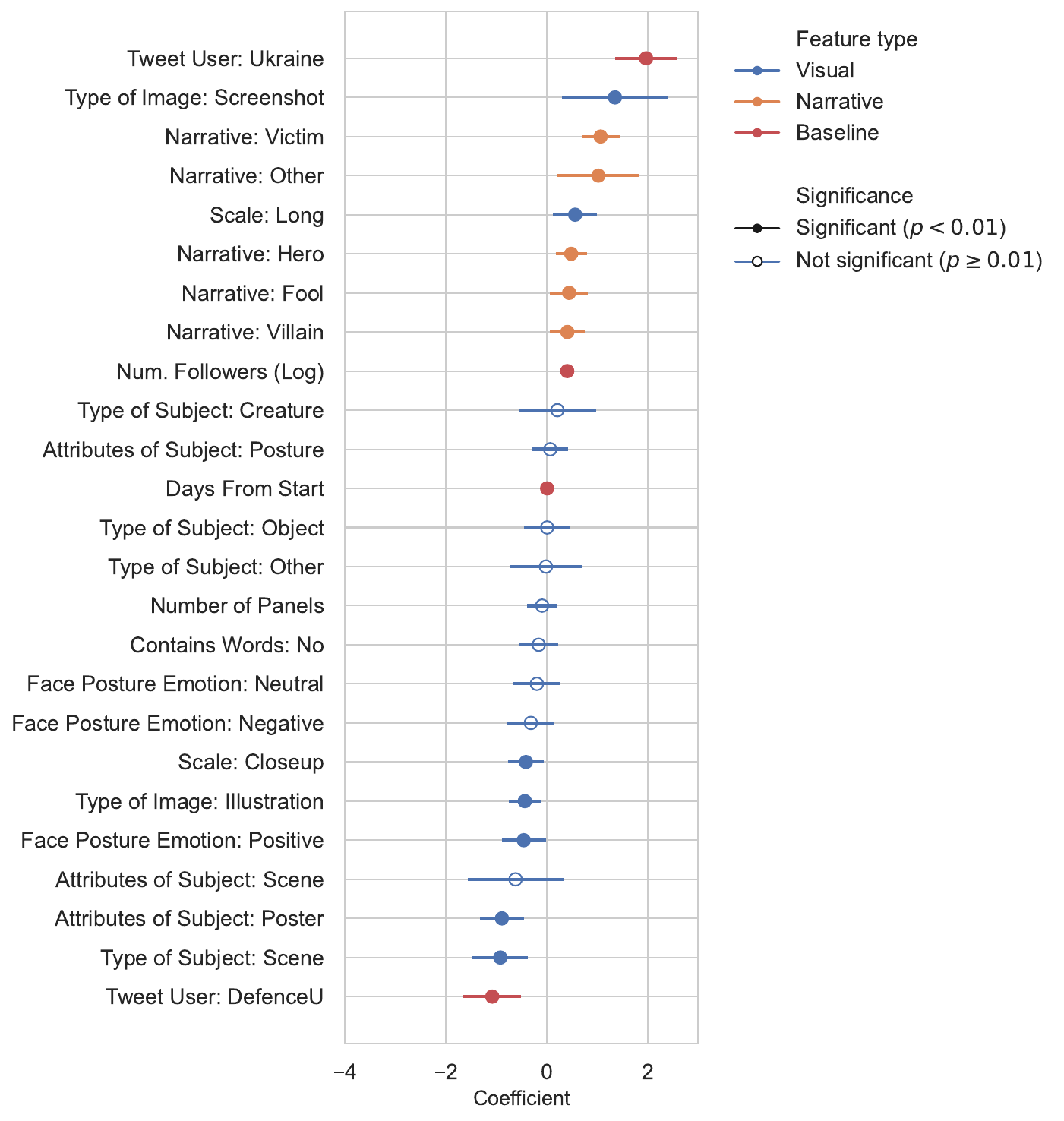} 
\caption{Coefficients and 95\% confidence intervals for the best-performing regression model of the popularity of a meme. The target variable is the $\log_2$ of the number of retweets: a coefficient of $+1$ indicates that a feature is associated with double the amount of retweets.}
\label{fig:best-model-coefficients}
\vspace{-.5\baselineskip}
\end{figure}

We now move to modeling the popularity of the content shared by these accounts.
We use a linear regression with the content variables outlined above as independent variables.
Categorical variables are represented via one-hot encoding with ``none'' as the reference case.
The target variable---the number of retweets---is transformed via log base 2, such that a coefficient of 1 indicates the number of retweets doubles, while a -1 that it halves.

First, we investigate which types of features are more effective in predicting the success of a tweet.
\Cref{tab:model_improvements} shows the performance of models with different subsets of explanatory variables.
All the models include the `baseline' variables: the identity of the author, the number of their followers, and the number of days since the invasion.
The other models use the variables grouped as visual, actor, emotion, intent, and narrative.
We perform model selection based on the BIC.
The best-performing model uses visual and narrative variables, and achieves $-43.9$ $\Delta$BIC compared to the baseline, with an Adjusted $R^2$ of 0.622.
Remarkably, the Narrative model alone (with only 5 features) performs almost as well as the Visual model (state of the art with 16 features) in terms of Adjusted $R^2$, and even better in terms of $\Delta$BIC.
This result demonstrates the importance of narratives in determining content popularity.

\Cref{fig:best-model-coefficients} shows the coefficients for the best model and their \rev{95\%} confidence intervals.
As the non-governmental \umf account is treated as a ``base case'', we find \ukr to have comparatively more retweets and \dou comparatively fewer. 
We have also computed a similar model in which, instead of using indicator variables for each account, we use a dummy variable for whether the author of the tweet is a government account (\ukr or \dou). 
In this model, this indicator variable is not significant (the effects of the two government accounts cancel each other out).
This result suggests that, at least for the present dataset, the popularity of the message is not necessarily tied to whether the account is governmental, but rather to the identity of the account and to the content itself.

Among the variables with the largest positive coefficients, we find the narrative-related variables, with the addition of screenshot types of images and long-scale images. 
The fact that all narrative coefficients are positive implies that the very existence of a narrative is associated with greater engagement in terms of retweets.
Out of these narratives, the \emph{victim} one is the most resonant: its presence is associated with an increase of \IncreaseNarrativeVictim in the number of retweets.
The other benevolent narrative, \emph{hero}, is associated with an increase of \IncreaseNarrativeHero.
The two narratives focusing on the malevolence of the enemy are instead less powerful (\IncreaseNarrativeFool for \emph{fool} and \IncreaseNarrativeVillain for \emph{villain} respectively), but still positive compared to the absence of a narrative.

Further, less popular content includes posters and illustrations, as well as those of scenes (as opposed to people or objects).
However, one visual variable that is positively associated with retweets is whether it is a screenshot: these are often calls for donations and screenshots of news.

Overall, we find narratives to be a powerful predictor of the amount of attention a piece of content receives, while emotional and actor variables have a smaller impact.

\subsection{Audience characterization}
\label{sec:audience}

As summarized by the words of the creator of \ukr reported in \Cref{sec:intro}, one of the main goals of Ukrainian memetic warfare is to reach ``large and distant target audiences'', and thus influence global public opinion.
In this section, we study which countries are more receptive to Ukrainian messages, and how this receptiveness relates to narratives and to each country's actions.

To examine the variability in responses in different countries, we look at retweets and geo-locate their authors to a country (see \Cref{sec:method}).
We normalize the number of retweets in a given country by its population size, thus obtaining a measure of retweets per capita.
This measure is a proxy for the amount of interest in Ukrainian tweets in a given country from the general population.
Note that we choose not to adjust the measure by Twitter or Internet usage within the country: we do not wish to equalize countries where Twitter is more likely to be read with those where it is not.
Indeed, Twitter usage is a mediator between the effect of Twitter messages and the support in the general population for sending aid to Ukraine, \rev{i.e., the larger the penetration of Twitter, \rvf{the more chances people in that country might be exposed to Twitter propaganda messages, all other factors being equal}.}
Therefore it would be incorrect to include it as a control variable in this case~\rev{\citep{pearl2012causal,pearl2013linear}}.
Finally, to measure the support of a country to Ukraine in the conflict, we use economic and military assistance normalized by GDP, as explained in \Cref{sec:method}.

\begin{figure}[t]  %
\centering
\includegraphics[width=0.9\textwidth]{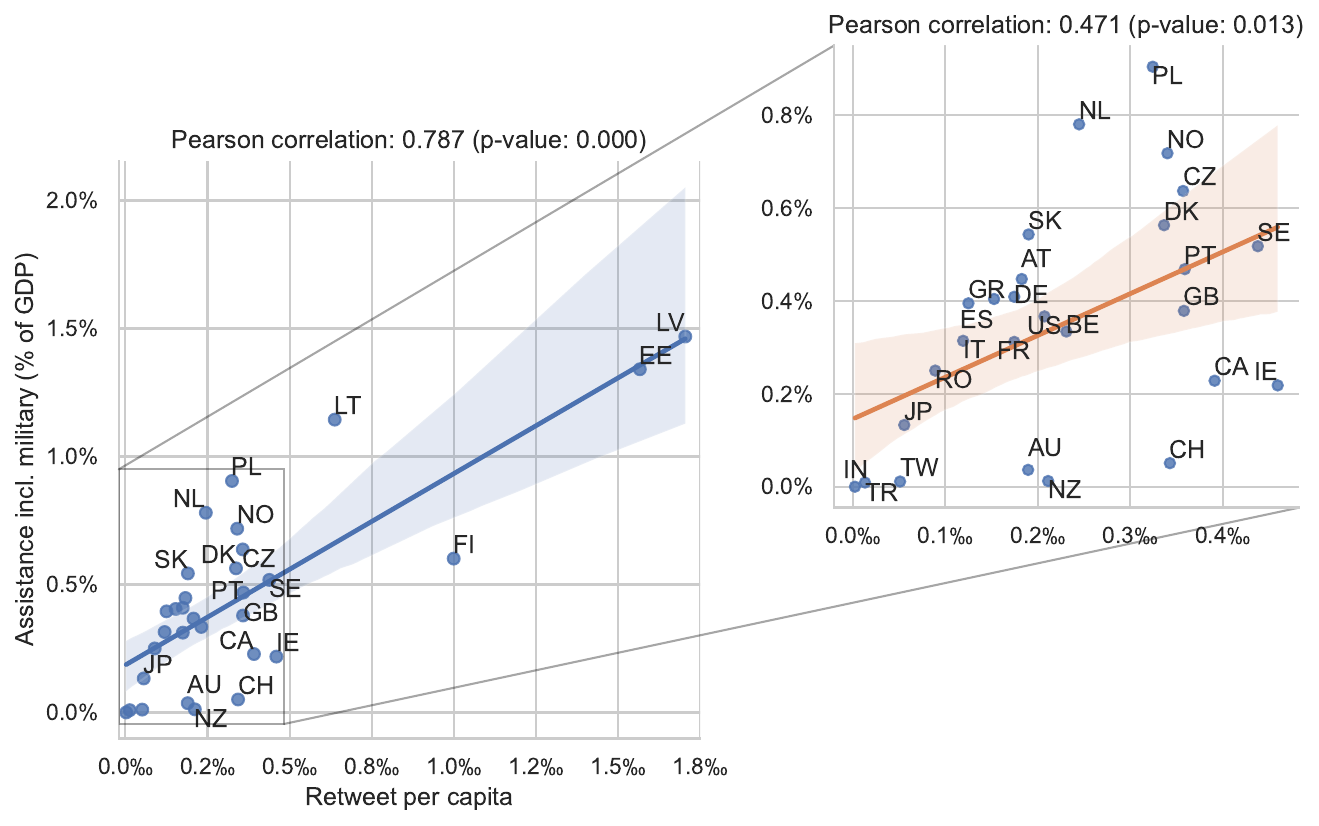} %
\caption{Number of retweets normalized by population, versus total bilateral and EU assistance to Ukraine, by country. Left: all countries. Right: excluding outliers. Both the regression including outliers (blue) and the one excluding them (orange) are significant at $p<0.05$. The shaded area represents the 95\% CI.}
\label{fig:retweetsPerCapitaVsAssistance}
\vspace{-.5\baselineskip}
\end{figure}

\Cref{fig:retweetsPerCapitaVsAssistance} shows these two measures, retweet per capita and assistance, for each country present in our dataset.
First, countries bordering Russia have a high retweet rate per capita (above 0.5\textperthousand): their attention to Ukrainian tweets is very high.
In particular, Baltic countries bordering Russia---Latvia, Lithuania, and Estonia---are outliers both in terms of assistance to Ukraine and of engagement.
Finland, also bordering Russia, presents a high retweet rate, albeit with a lower aid statistic. %
Other European countries---northern (Norway, Denmark), eastern (Czech Republic, Poland), and English-speaking ones (Ireland, United Kingdom)---have high rates of retweets per capita (above 0.3\textperthousand).

The scatterplot shows a strong correlation between the two variables: with a Pearson coefficient of $\rho = 0.787$, attention to Ukrainian content on Twitter is related to the economic and military assistance that the country provides to Ukraine.
While European countries bordering Russia are outliers, the correlation after excluding these countries is still a reasonably strong $\rho = 0.471$ ($p<0.05$, inset on the right).
Thus, the relationship extends beyond the countries bordering Russia.
To check for robustness, we also perform this analysis disaggregating by account, and the results are in line with those shown in \Cref{fig:retweetsPerCapitaVsAssistance} (coefficients $> 0.7$ and p-values $< 10^{-4}$).
In other words, picking any subset of the accounts would not change this result, which therefore seems more general than our particular choice of accounts.
Conversely, the same analysis using the Eurobarometer survey shows no significant correlation between retweets per capita and popular support for financial aid to the Ukrainian military effort: the correlation we observe between the effectiveness of Ukrainian content on Twitter and the action of governments does not translate when considering support from the general population.

\begin{figure}[t]  %
\centering
\subfloat[Retweet per capita]{\includegraphics[width=0.2\textwidth]{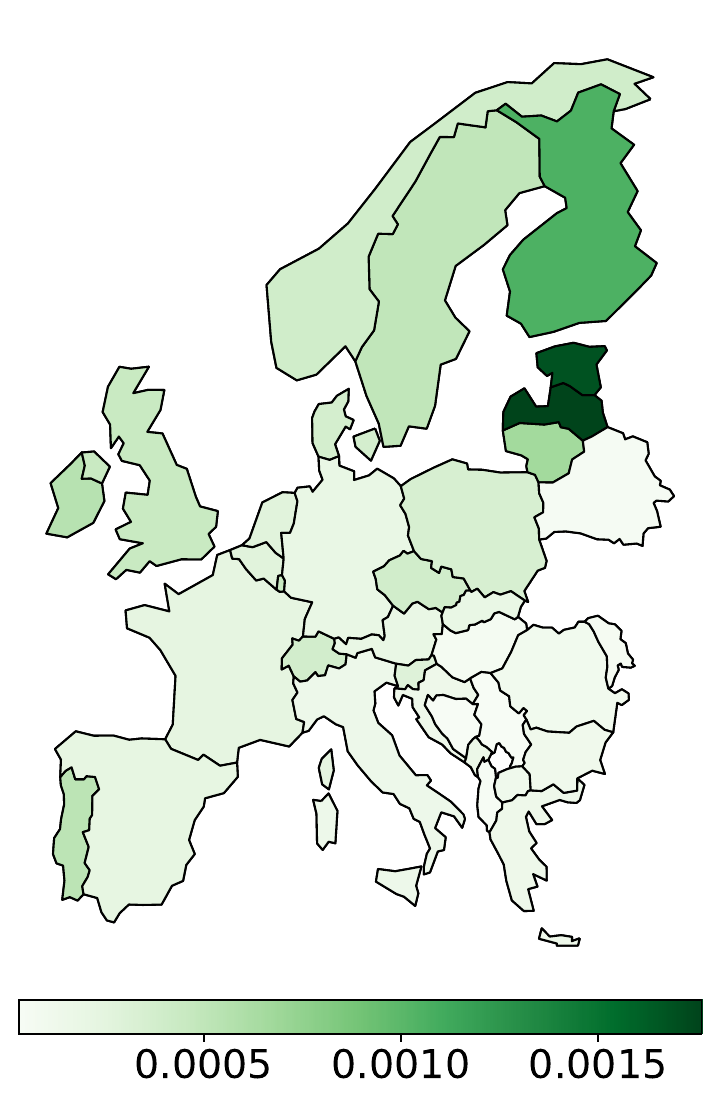}}
\subfloat[Log odds: Hero]{\includegraphics[width=0.2\textwidth]{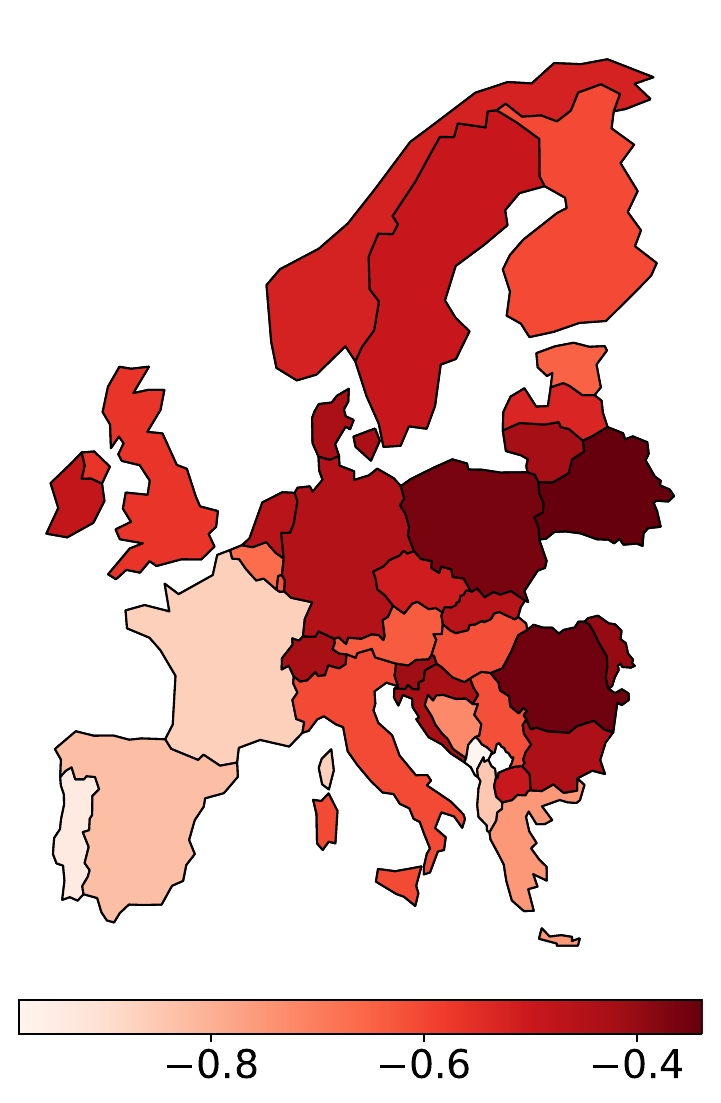}}
\subfloat[Log odds: Victim]{\includegraphics[width=0.2\textwidth]{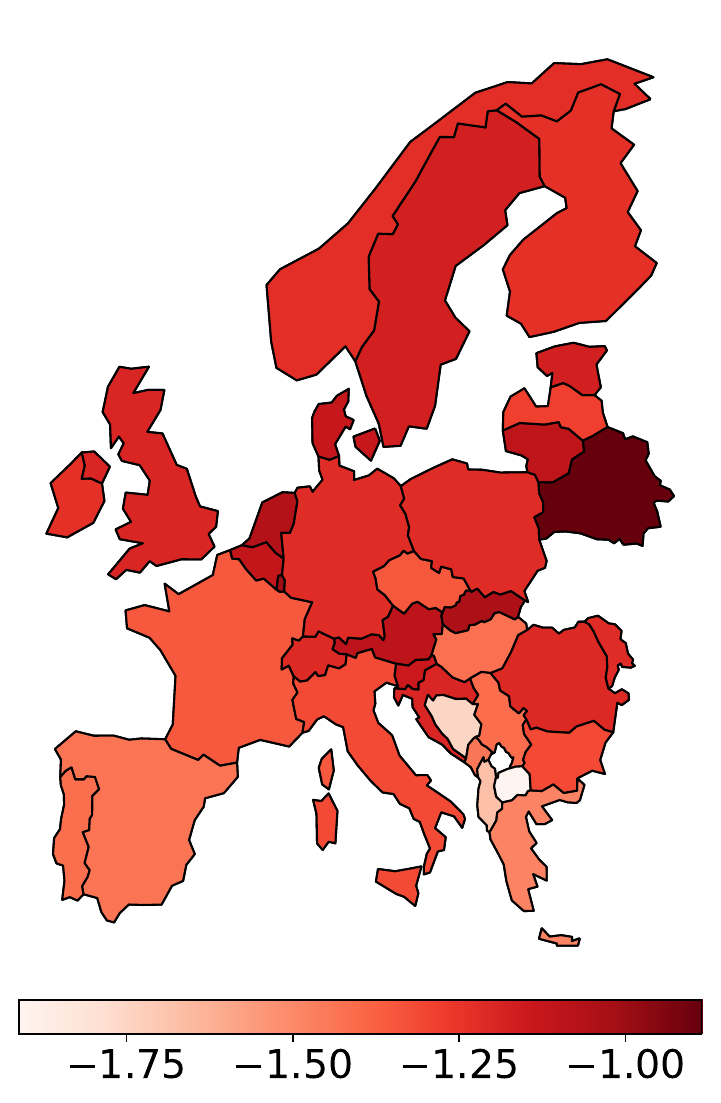}}
\subfloat[Log odds: Villain]{\includegraphics[width=0.2\textwidth]{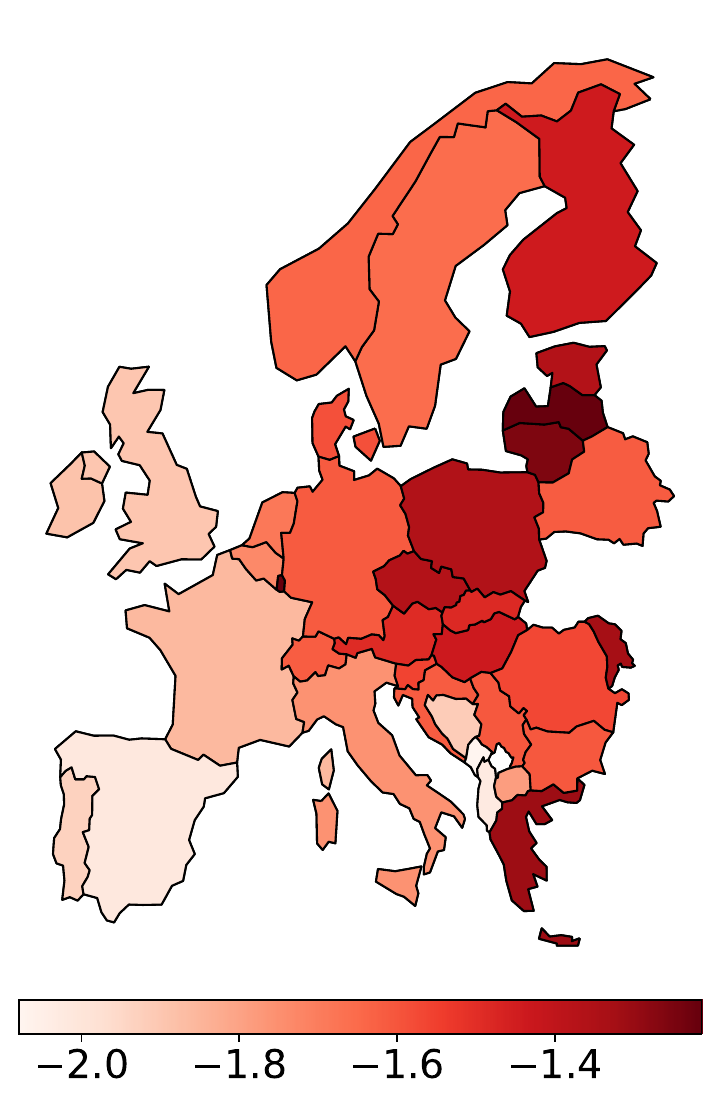}}
\subfloat[Log odds: Fool]{\includegraphics[width=0.2\textwidth]{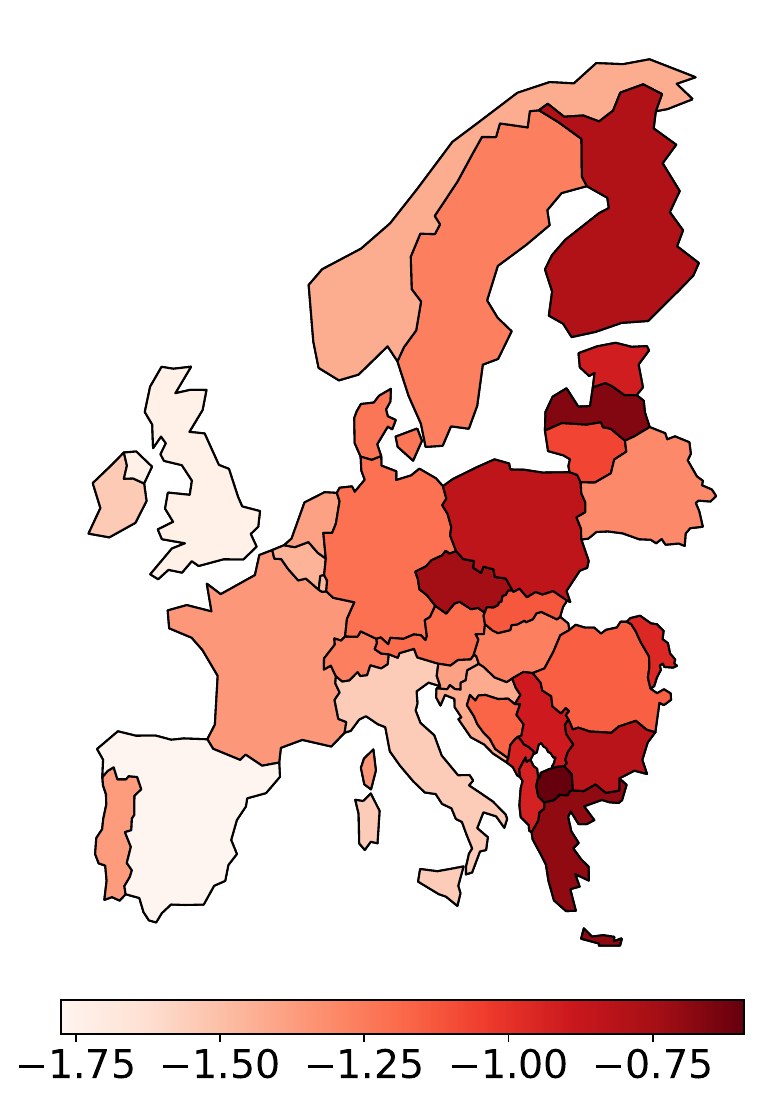}}
\caption{Geographic distribution of retweets. Left: retweet rate per capita in a country. Rest: log odds of retweeting content with a specific narrative.}
\label{fig:maps}
\vspace{-.5\baselineskip}
\end{figure}

Let us now break down the geographical resonance of each narrative.
To this end, we compute the odds of users in a country to retweet content with a particular narrative.
An important working assumption in this analysis is that geolocation is not biased compared to the narrative that is shared, i.e., there is no reason to believe that a person who geolocates themself prefers one narrative over another. %
\Cref{fig:maps} shows the retweet rate and the per-narrative log odds for each country in Europe.
The focus on Europe derives from it being the geographic location of the conflict, and because it is the continent with the highest retweet rate per capita (0.3\textperthousand, compared to 0.09\textperthousand \xspace in North America, which is second). %
The leftmost map shows the retweet rate normalized by population, which shows again the outsized attention by the Baltic countries and Finland.
The retweets by narrative (rest of the plots) often present a gradient from West to East: countries closer to Russia are more likely to retweet posts with a discernible narrative. 
There are, however, some differences across narratives: the `victim' narrative presents a more uniform geographical spread.
In contrast, countries bordering Russia and Ukraine---such as Finland, Poland, and the Baltics---show a stronger resonance with the `villain' narrative focusing on the Russian threat.
Baltic states are particularly inclined to amplify the archetypes of villain and fool, interestingly together with Greece (we may speculate that historical parallels with Cyprus' experience of the contested occupation by Turkey may provoke attention from Greece).
The `hero' narrative appeals both to some Western countries---such as the United Kingdom, a strong supporter of Ukraine in the war---as well as those closer to the conflict, including Belarus.
Note that Belarus has strong regulations on social media access, thus those who express themselves on social media are likely to be either those in opposition to the government and dedicated enough to use a VPN, or those who are allowed by the government on the site.\footnote{\url{https://www.cyberghostvpn.com/en_US/privacyhub/countries-ban-social-media}}

\begin{figure}[t]  %
	\centering
	\includegraphics[width=0.75\textwidth]{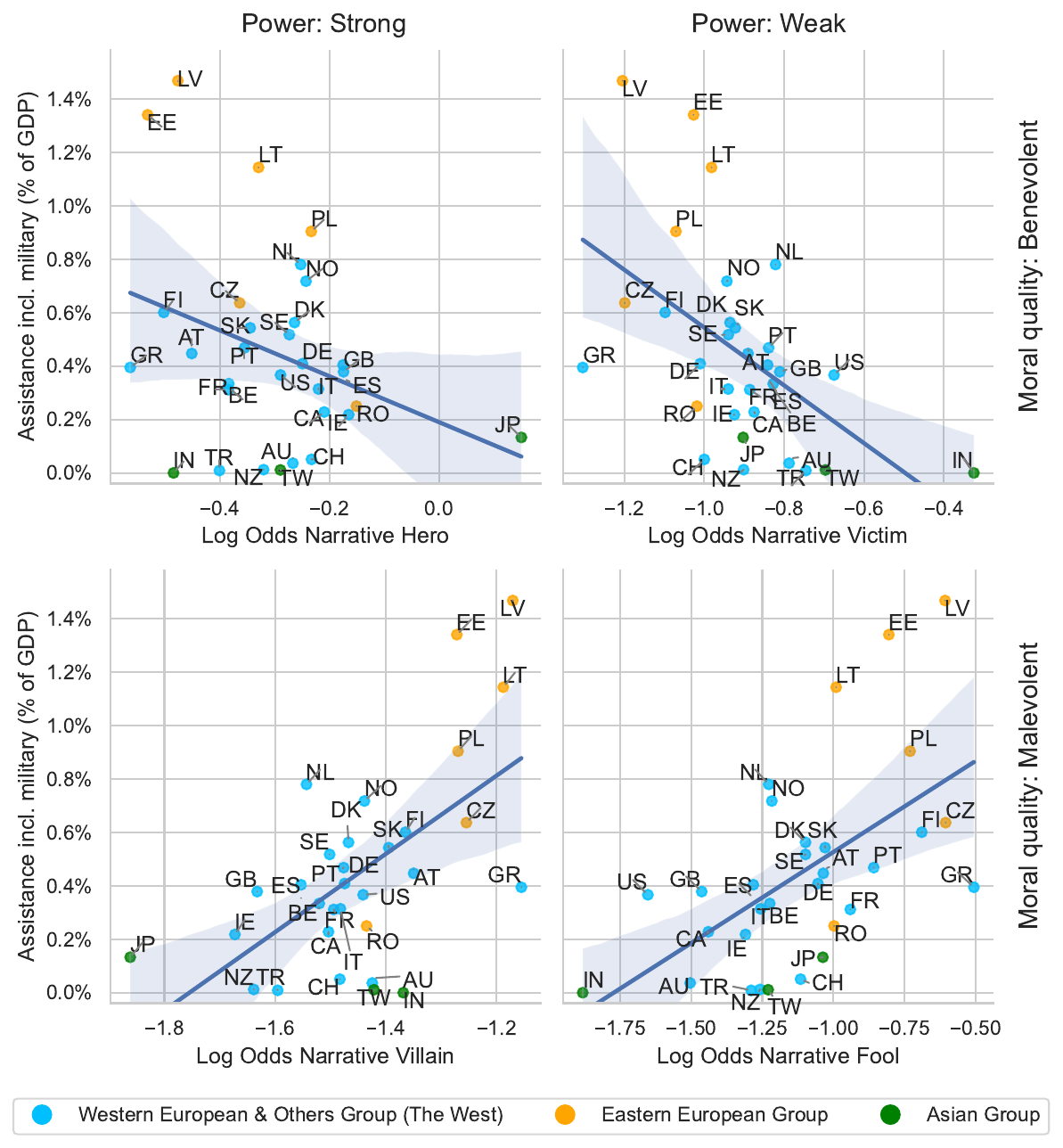}
	\caption{Log odds of retweeting content with a particular narrative, versus total bilateral and EU assistance to Ukraine, by country.
	Colors indicate the United Nations Regional Group of each country. The shaded area represents the 95\% CI.
	The columns indicate the power quality of the narrative, while rows indicate their moral quality.
	The assistance is driven mostly by military aid, whereas non-military aid has no significant correlations with narrative preference. 
	Malevolent narratives that target the malevolent moral quality of the enemy are positively correlated with assistance.}
	\label{fig:LogOddsNarrativeVsAssistance}
	\vspace{-.5\baselineskip}
\end{figure}

Finally, we investigate how the receptiveness to different narratives is related to the actions of a country's government in the conflict.
\Cref{fig:LogOddsNarrativeVsAssistance} shows the relationship between these log odds and the total assistance to Ukraine.
The total aid provided by each country to Ukraine, normalized by GDP, is positively associated with the spread of malevolent narratives (villain) and negatively associated with benevolent ones (hero).
We perform the same analysis also disaggregating by account, to evaluate which findings are account-specific and which are robust to our particular choice of accounts to study.
In doing so, we wish to consider also the levels of uncertainty of each log odds ratio (i.e., our measure of narrative effectiveness) in each country: a low number of retweets in a country implies that the associated odds ratio is uncertain.
Therefore, we use a Bayesian approach through a Monte Carlo simulation with 100k samples, wherein the probability of a retweet is modeled as a Beta distribution with parameters taken from the data (number of retweets with a given narrative vs without that narrative).
\Cref{tab:logodds_correlation} reports the results of this analysis.
While the results for some narratives are indeed account-specific, we find that the association between the effectiveness of the \emph{villain} narrative and financial assistance to Ukraine is robust to the choice of the account.
This finding seems to indicate that the focus on the evil character of the enemy is the most powerful narrative with governments that provide material assistance.
Moreover, we find that these effects are mostly due to military aid, while humanitarian aid has no statistical relation to the resonance of narratives in our dataset (plots omitted for brevity).
The Eastern European countries (closest to the conflict) both contribute the most aid (proportional to their GDP) and prefer the villain narrative. 
Interestingly, the second most stable relationship is a negative one between the preference for the hero narrative and aid: the countries that contribute the most aid do not show a preference for content portraying Ukraine as a hero.
Again, a similar analysis shows no significant results for the Eurobarometer survey.

\begin{table}[htb]
    \centering
    \sisetup{table-number-alignment=right}
    \caption{Pearson correlation between the log odds of retweeting content with a particular narrative versus total bilateral and EU assistance to Ukraine (only values significant at $\alpha=5\%$ are reported), combined for all accounts, and for each account separately, as well as the total number of retweets involved in the computation (last column).
    These values account for the uncertainty in the estimates of the log odds of each country via a Monte Carlo approach with 100k samples, where the probability of retweet is modeled as a Beta distribution with parameters taken from the data.}
    \label{tab:logodds_correlation}
    \begin{tabular}{lSSSSS[table-format=6.0]}
    \toprule
                 & {Hero} & {Victim} & {Villain} & {Fool} & {Total Geo. RTs} \\
    \midrule 
        Combined & -0.312 & -0.503 & 0.573 &  0.561 & 224225 \\
        \ukr     &        & -0.344 & 0.281 &  0.375 & 70312  \\
        \dou     & -0.400 &  0.328 & 0.403 &        & 91627  \\
        \umf     & -0.400 &  0.613 & 0.430 & -0.362 & 62286  \\
    \bottomrule
    \end{tabular}
\end{table}

\section{Discussion}
\label{sec:discussion}

\rev{The results of this study have important implications for our understanding of online communication during war, especially within the theoretical frameworks outlined in \Cref{sec:relwork}.
We observed the international reach of the content produced by Ukrainian sources, particularly in reaching Western countries where Twitter's penetration is considerable.
This reach is particularly noteworthy since exposure to political memes is known to be a mediator for political involvement~\cite{ahmed2024breaking}. %
}
Ukrainian war memes that contain clear narratives prove to be particularly popular, and in particular, tweets with a victim narrative receive twice the number of retweets.
As social media is increasingly used to mobilize support~\cite{ye2023online,chapman2022rage}, narrative emphasis may be an important variable to control in communication campaigns.
Interestingly, the overall audience preference for the victim narrative captured here does not translate into action.
Instead, countries that contribute more to the Ukrainian war effort resonate more with the villain narrative.
We can speculate that countries united by the common fear of a powerful enemy are more willing to make sacrifices to help in the conflict.
Indeed, the countries that relate more to depictions of Russia as a grave threat are the ones closer to the conflict, such as the Baltic countries, given their shared history with and physical proximity to Russia.
For instance, soon after the invasion, Finland reversed a long-time policy of neutrality and applied for membership to NATO on 18 May 2022, which was granted on 4 April 2023.
In this sense, the propagation of the content posted by the examined Ukrainian accounts reflects politically relevant signals. 
\rev{At the same time, we find that while economic assistance in the military conflict is correlated with the spread of Ukrainian content on Twitter, it is not correlated with the popular support as measured by surveys (see \Cref{sec:audience}).
This unexpected result could hint at governments and engaged Twitter users being more in tune with each other than with the general population, and highlights the important role played by social media as a mediator for political activity.
}
Whether a government's decision to support another country monetarily is preceded by social media activity, especially involving state actor accounts, is a fascinating future work direction.
In summary, messages focusing on the Ukrainian victims are more effective in reaching a larger audience; but messages focusing on the danger posed by the enemy resonate more with countries willing to offer material assistance in the war. %

\rev{Our findings suggest several broader considerations on memetic warfare from the perspective of the general public.
\rvf{First of all, since meme creators are likely to be aware of the engagement created by victim narratives, it is important that audiences are able to respond knowledgeably.} %
As such, it would be helpful to increase public awareness of such mechanisms to improve digital media literacy. 
Currently, the EU's Digital Skills~\cite{vuorikari2022measuring} include the skill to ``critically evaluate the credibility and reliability'' of information.
We argue that evaluating the \emph{intent} behind the posting of content should also be taken into account.
Beyond individual literacy, reported grievances from warring governments should be verified by independent sources, such as supernational bodies and humanitarian organizations.
As noted by \citet{nyhan2021backfire}, in fact, while the efficacy of post-hoc fact checking is severely limited, it is important to prevent and disrupt the formation of linkages between group identities and false claims.
In this regard, particular attention is needed towards misusing imagery or taking it out of context, a common misinformation tactic that adds to the perception of authenticity~\cite{fazio2020outofcontext}.}

\rev{Another important observation is that messages hinting at the pure evilness of the enemy---e.g., portraying its soldiers as stealing toys from dead children---are particularly likely to spread in countries that are more involved in the war through economic and military support.
While we cannot establish the direction of the causal arrow in this case---i.e., we do not know if involved countries are a more fertile ground for these messages, or if such messages encourage involvement in the war---this finding suggests memes might reflect the broader trend of increasing geopolitical tension.
Group identification and unempathetic psychological responses toward the rival out-group have been found to drive media consumption and engagement~\cite{wakefield2023intergroup}.
Negative affect (which subsumes anger) has been recognized as a powerful mediator between ``counter-empathy'' and social media engagement~\cite{weiss2019clicking}.
As such, a high level of alert is particularly needed regarding the possibility of unsubstantiated wartime ``atrocity tales''~\citep{bromley1979atrocity} in the countries involved in a conflict, lest they provoke the opposite reactions of either inciting the dehumanization of entire populations, or the desensitization of the public. %
There are historical precedents for this scenario:
during World War I, all participating nations widely employed atrocity propaganda, which significantly contributed to fostering the surge of nationalism during the escalation of the war into a global conflict~\cite{cull2003propaganda}.
At the same time, such usage contributed to a growing discrediting towards reports of atrocities, that led to their diminished propagation during World War II~\cite{taylor2013munitions}. 
}

\rvf{These narrative elements have been previously well recognized as key in the media war.
Other analyses of the Ukrainian war~\cite{moutzouridis2023war} also recognized the frames of good vs.~evil and us vs.~them as essential in understanding the narratives employed by both sides.
The focus on the portrayal of the enemy in a `Cold-War' style was already noted by~\citet{boyte2017analysis} in \citeyear{boyte2017analysis}.
In the Syrian War, while the evil vs. good framing was also employed by traditional media, %
\citet{cole2022syrian} noted also how the co-presence of different state-sponsored propaganda actors on social media created a confusing and fragmented information environment.\footnote{\url{https://www.theatlantic.com/international/archive/2018/03/cnn-effect-syria/554387}}}
\rev{That said, not all usage of narrative-laden messaging campaigns on social media is necessarily problematic.
Indeed, social media has been used to build collective identities through narratives, which has been widely studied in the field of human-computer interaction.}
For instance, \citet{das2022collaborative} postulate that using a social media platform as a space to address colonial grievances and build a self-concept fosters a ``narrative resilience'', which serves as a ``collective and reflexive mechanism through which people work to generate resilience''.
During the annotation process, we encountered many examples of \citet{laenui2000processes}'s decolonization phases, including mourning, commitment, and calls to action, and yet other emphases may be present as the conflict develops and hopefully ends (although applying the decolonization lens to the Russo-Ukrainian conflict is controversial).
Further, narrative building has been used to engage marginalized communities~\cite{pei2022narrativity}, strengthen their voice~\cite{halperin2023probing}, and promote conflict reconciliation~\cite{stock2008co}---tasks that may become important towards the end of the conflict.
Recently, social media narratives have been shown to provide Ukrainian refugees with a way to maintain an ``emotional, dynamic, and constantly updating'' bond with their homeland, and a space for the ``negotiation and performance of ethnic and national identities''~\cite{kozachenko2021transformed}.
Given the importance of narratives found in this study, examining their role in establishing the self-concept of Ukraine as an independent state, and its relationship with Russia, the West, and the rest of the world, is an important future research direction.

Another crucial actor impacting the spread of propaganda is the platform, \rvf{its algorithms,} and its moderation policies.
\rvf{
The spreading of memes on social media is largely influenced by platform-specific algorithms, which curate content based on user interactions and engagement patterns.
It is well-known that content with negative sentiment attracts more engagement from users on social media~\citep{Rathjeanimosity,BradyEmotion,TsugawaNegative,hansen2011good}.
These engagement signals are used by the platform algorithms to decide which content to promote: this algorithmic amplification ultimately determines what becomes viral~\citep{dujeancourt2023effects}.
Some content may achieve virality across multiple platforms, creating a cross-platform dynamic in which memes spread through different online ecosystems, influenced by different algorithms and user demographics~\cite{shifman2013memes}.
This self-reinforcing loop not only accelerates the spread of memes~\cite{ling2021dissecting}, but also shapes the cultural relevance of the memes themselves, as they evolve and adapt to the affordances of each platform~\cite{nahon2013going}.
At the same time, states and other institutional actors can mobilize their social, economic, and cultural capital to meddle in the work of algorithms: as such they have the means to further their long-term strategic vision~\cite{bonini2024algorithms}.
}
Shifting our attention to moderation policies, we did not find any evidence of moderation of the content posted by the examined accounts. 
However, platforms have been known to moderate political speech. 
It has been revealed that Facebook maintains a blacklist of ``organizations with a record of terrorist or violent criminal activity'',\footnote{\url{https://theintercept.com/2021/10/12/facebook-secret-blacklist-dangerous}} whereas Twitter has been known to ``shadow ban'' users by limiting the propagation of their content.\footnote{\url{https://mashable.com/article/twitter-visibility-limited-labels-roll-out}} 
The application of moderation actions, especially when applied to political speech, is often controversial.
In this sense, platforms' choices can shape public discourse: this selective moderation, or lack thereof, is a political choice that warrants critical reflection, especially in the context of memetic warfare.
\rvf{The platform's behavior is a black box: the algorithms are proprietary to the respective platforms and are not available for direct scrutiny, and the moderation decisions are made by in-house teams, possibly assisted by more algorithms.}
Barring voluntary greater transparency by the platforms, systematic auditing of the political speech and its reach may be the only way to provide a clearer view of the role platforms play in the spread (or limitation) of propaganda.

\spara{\rev{Limitations.}} The insights of this study are limited by its unique setting: the period of the Russian invasion, the selection of the accounts, the peculiar affordances the Twitter platform presents, and the uneven adoption of the platform around the world.
Nevertheless, our finding that users from countries that support Ukraine the most also retweet the content from major Ukrainian accounts at a higher rate suggests a synchrony between the supportive actions of Twitter users (which could be considered to be a form of slacktivism~\cite{lee2013does}) and concrete financial decisions by the governments of their countries.
Conversely, the responses to the Eurobarometer survey question on sending help to Ukraine had no significant correlation with the propagation of the Ukrainian tweets, which suggests that users on Twitter are more ``in tune'' with the decisions of their governments than a representative population sample. %
It is possible that the retweeters include government entities and actors, as well as automated accounts.
Unfortunately, we are unable to apply the standard bot detection techniques~\cite{davis2016botornot} as the platform closed down its API in early 2023. 
However, the effect of such accounts should be limited in a retweet analysis, as each account may retweet a post once at most.
\rev{Our methodology focuses on the popularity of content posted by official accounts.
However, the propagation of memes often takes place organically, outside of the retweet mechanism, potentially spreading to other communities. 
To study the larger evolution of the memes captured in this study, the content and its sources and derivatives should be tracked across accounts, and, ideally, across platforms.
Especially since it has been shown by \citet{zannettou2018origins} that memes often originate in fringe communities, capturing the evolution of the narratives within such communities (e.g., conspiracy theories) and their interaction with propaganda would be an exciting future research direction.
}
Finally, although we strove to identify the narratives from the points of view of the posters, some subjective labels %
might have been affected by the opinions of the authors (see Positionality Statement below). 
We hope that sharing this data will allow for reproducibility and an easier re-examination of the narrative extraction process.

As the content analyzed here was produced during wartime, special care must be taken in an ethical analysis of the potentially sensitive material it contains, despite being published by verified accounts.
For instance, \dou posts a variety of reporting from the battlefields, including active war situations involving the use of artillery, images of combatants, and distressing images of affected civilians. 
We assume that the posting accounts have obtained sufficient permissions to post images of individuals, and have performed intelligence analysis to make sure no important information is being disclosed. 
Nonetheless, in this manuscript, we endeavored not to use any specific names or locations.
Further, by its very nature, this data contains images of and information on vulnerable populations (including children), some of which are affected by the conflict. 
Special care must be taken that their images and other contexts are not misused and that any harm is minimized in subsequent analyses.

\subsection*{Positionality Statement}

All authors reside in ``the West'', and one was born in Russia. 
Although it is our aim to conduct the analysis as objectively as possible, this positionality is likely to affect some interpretation of the results. 
We make the data and annotations available for the examination of the research community, and possible alternative interpretations.

\bibliographystyle{ACM-Reference-Format}
\bibliography{references}

\clearpage
\appendix
\renewcommand\thefigure{\thesection.\arabic{figure}} 
\renewcommand\thetable{\thesection.\arabic{table}} 
\setcounter{figure}{0}
\setcounter{table}{0}

\section{Original data collection keywords}
\label{sup:keywords}

\begin{figure}[h]  %
\centering
\includegraphics[width=0.8\textwidth]{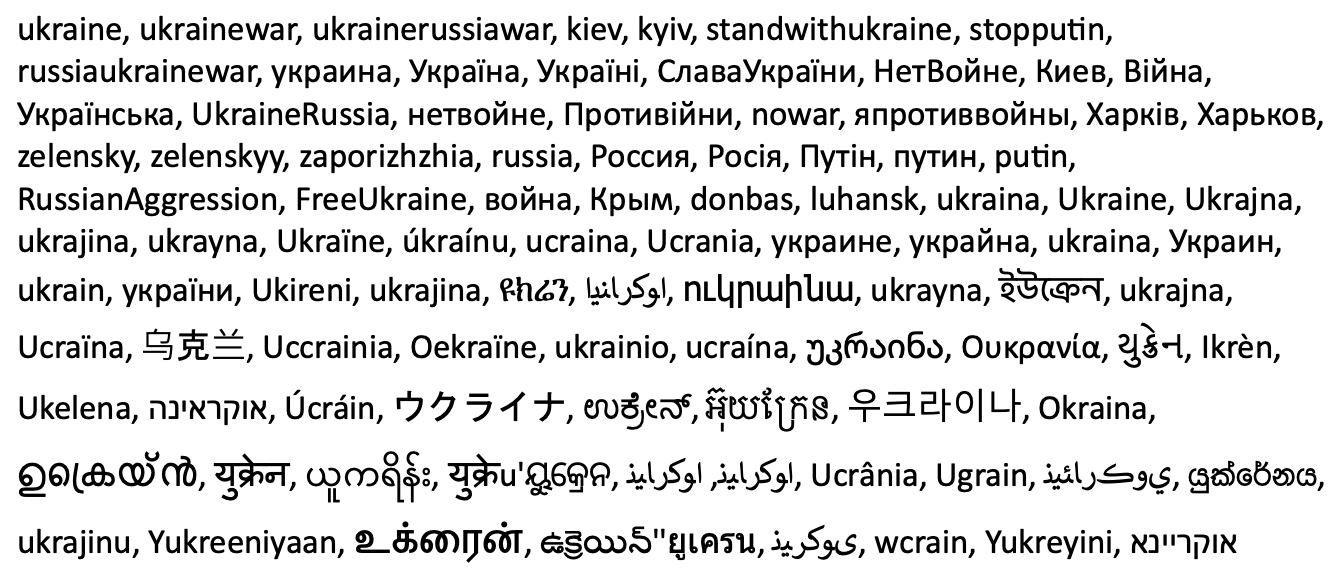}
\caption{Keywords of the large data collection used to identify top users}
\label{fig:keywords}
\end{figure}

The first data collection was performed using the Twitter Streaming API using keywords determined using snowball sampling during the first week of the invasion. The keyword ``Ukraine'' was translated into some of the most used languages and alphabets, as can be seen in \Cref{fig:keywords}. 

\section{Annotation codebook}
\label{sec:codebook}

\begin{enumerate}

 \item Number of panels. 
 \begin{itemize}
  \item Single panel: images that are composed of only one image
  \item Multiple panels: images/posts that are composed of a series of images
 \end{itemize}

 \item Type of image.
 \begin{itemize}
  \item Photo: a picture taken by a camera
  \item Screenshot: an image of a screenshot taken from a computer screen
  \item Illustration: a drawing, painting, or printed work of art
 \end{itemize}

 \item Scale. 
 \begin{itemize}
  \item Close-up: a shot that tightly frames a person or object, such as the subject’s face taking up the whole frame
  \item Medium shot: a shot that shows equality between subjects and background, such as when the shot is ``cutting the person in half''
  \item Long shot: a shot where the subject is no longer identifiable and the focus is on the larges scene rather than on one subject
  \item Other: there is no clear subject or background
 \end{itemize}

 \item Type of subject. 
 \begin{itemize}
  \item Object: refers to a material thing that can be seen and touched, like a table, a bottle, a building, or even a celestial body
  \item Character: refers to people or anthropomorphized creatures/objects, such as cartoon characters
  \item Scene: when the situation or activity depicted in an image is its main focus, instead of it being on the single characters or objects depicted in it
  \item Creature: refers to an animal that is not anthropomorphized
  \item Text: text is the focus of the image.
  \item Other: anything else, e.g. posters or infographics
 \end{itemize}

 \item Attributes of the subject. For images whose subject is one or more characters, we consider whether the image’s visual attraction lies with the character’s facial expression or with their posture. For the other attributes, we identify five features.
 \begin{itemize}
  \item Facial expression: only for a character
  \item Posture: only for a character
  \item Poster: informative large scale image including both textual and graphic elements. There are also posters only with either of these two elements. Posters are generally designed to be displayed at a long distance. It informs or instructs the viewer through text, symbols, graph, or a combination of these.
 \end{itemize}
 
 \item Contains words. Whether words are a part of the image, including superimposed on top of an existing image.

 \item Character face/posture emotion.
 \begin{itemize}
  \item Positive, Negative, or Neutral: only for a character annotated as facial expression
 \end{itemize}

 \item Narrative (possible more than one).
 \begin{itemize}
  \item Hero (benevolent, strong)
  \item Victim (benevolent, weak)
  \item Villain (malevolent, strong)
  \item Fool (malevolent, weak)
  \item Other
  \item None
 \end{itemize}

 \item Emotional appeal. [Open label]
 \begin{itemize}
  \item Humor: satirial, sarcastic, or funny
  \item Fear: threats, harms, or calamities 
  \item Outrage: scandals, corruption, or moral hazards
  \item Pride: objects/scenes of adoration or affection
  \item Compassion: sympathy and sorrow for another who is stricken by misfortune
  \item None
 \end{itemize}

 \item Intent. [Open label]
 \begin{itemize}
  \item Informational
  \item Call to action
  \item Other
 \end{itemize}
 
 \item Actor. [Open label]
\end{enumerate}

\section{Inter-annotator agreement}
\label{sec:agreement}

\Cref{tab:agreement} shows the inter-annotator agreement between the authors, measured by Krippendorff's alpha, for the annotation of the data described in Methods.
In aggregate, the labelers are mostly in agreement apart from the open label of Actor (which was later cleaned and aggregated).

\rev{
\begin{table}
    \sisetup{
        detect-mode,
        table-align-text-pre    = true,
        round-mode              = places,
        round-precision         = 3,
        table-format		= 1.3
        }
\caption{\rev{Inter-annotator agreement (Krippendorff's alpha).}}
\centering
\label{tab:agreement}
\begin{tabular}{lSSS}
\toprule
{Feature} & {\ukr} & {\umf} & {\dou} \\
\midrule
Number of panels & 1.000000 & 1.000000 & 0.825123 \\
Type of image & 0.753927 & 0.771525 & 1.000000 \\
Scale & 0.643154 & 0.401779 & 0.674757 \\
Type of subject & 0.737762 & 0.663866 & 0.845982 \\
Attributes of subject & 0.597015 & 0.630600 & 0.859589 \\
Contains words & 0.884521 & 0.847512 & 0.765306 \\
Character emotion & 0.357143 & 0.581773 & 0.691429 \\
\cmidrule(lr){1-4}
Actor (open) & 0.328729 & 0.307305 & 0.156863 \\
Emotional appeal (open) & 0.561658 & 0.195199 & 0.414651 \\
Intent (open) & 0.627483 & 0.788530 & 0.510000 \\
Narrative & 0.703857 & 0.590012 & 0.644272 \\
\bottomrule
\end{tabular}
\end{table}
}

\section{Visual content features}

\begin{figure}[h]  %
\vspace{2\baselineskip}
\centering
\subfloat[Number of panels]{\includegraphics[width=0.23\textwidth]{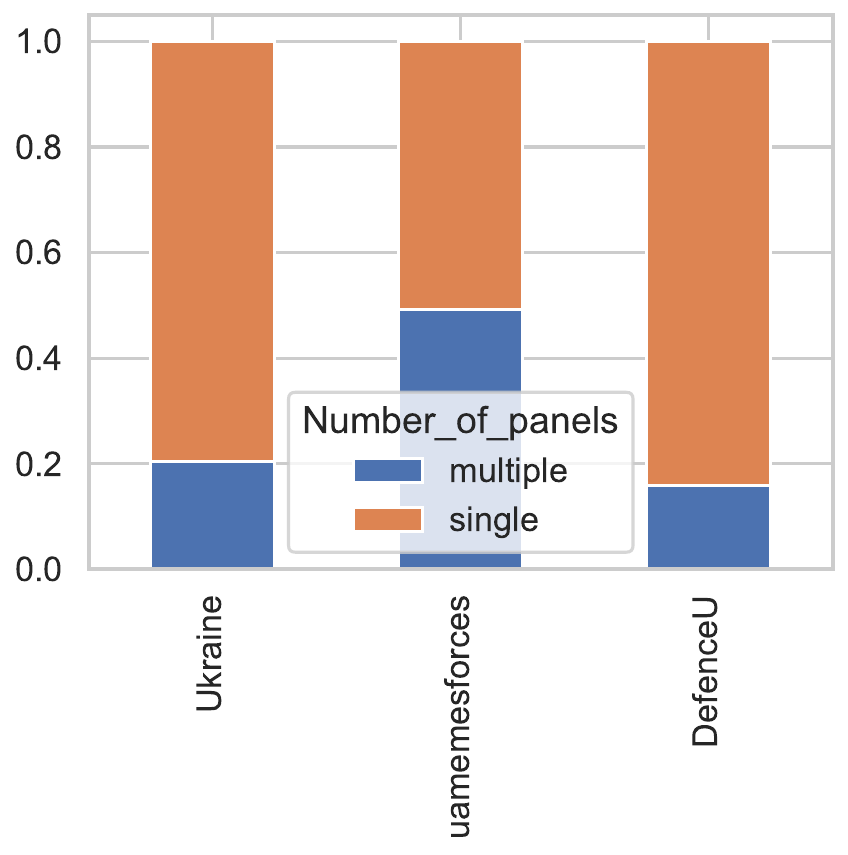}} 
\subfloat[Type of image]{\includegraphics[width=0.23\textwidth]{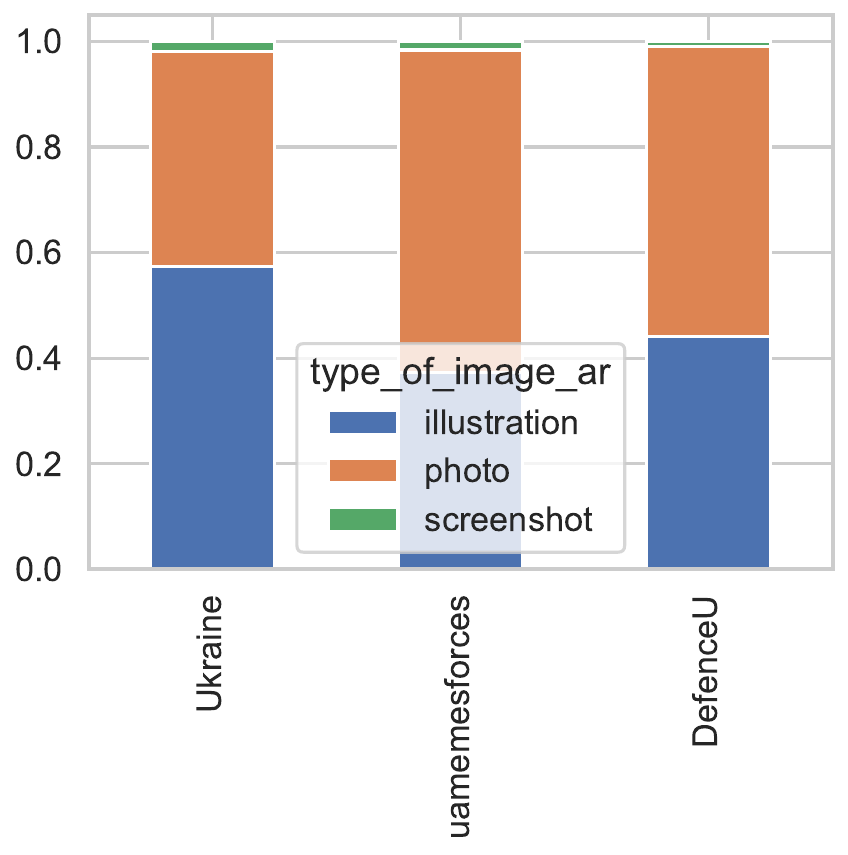}} 
\subfloat[Scale]{\includegraphics[width=0.23\textwidth]{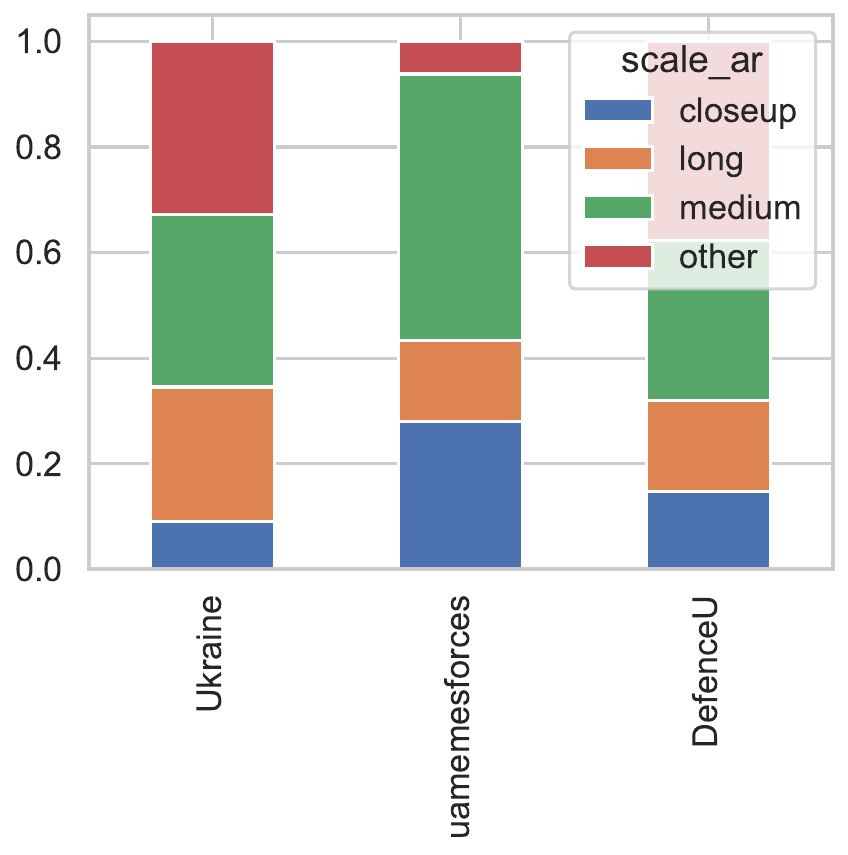}}  
\subfloat[Type of subject]{\includegraphics[width=0.23\textwidth]{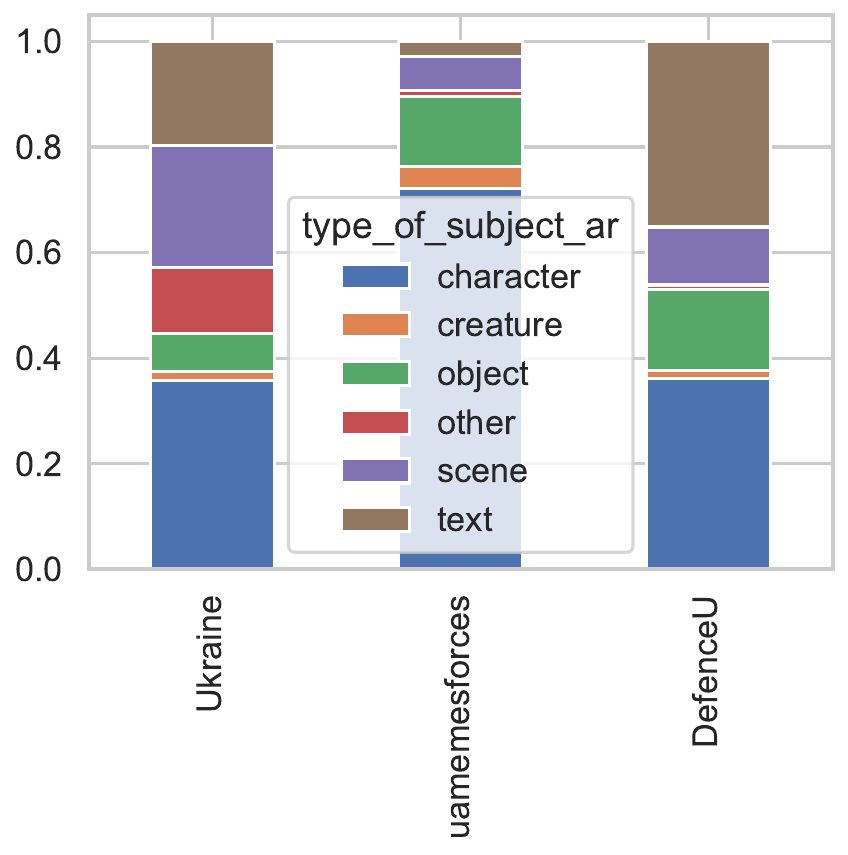}} \\
\subfloat[Attribute of subject]{\includegraphics[width=0.23\textwidth]{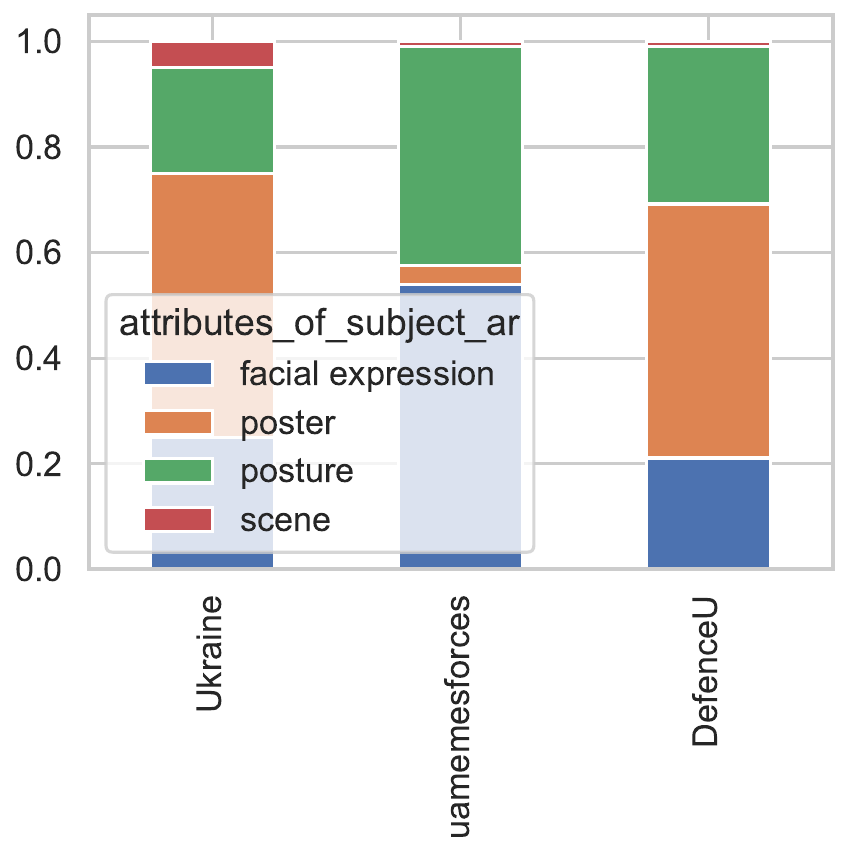}}  
\subfloat[Contains words]{\includegraphics[width=0.23\textwidth]{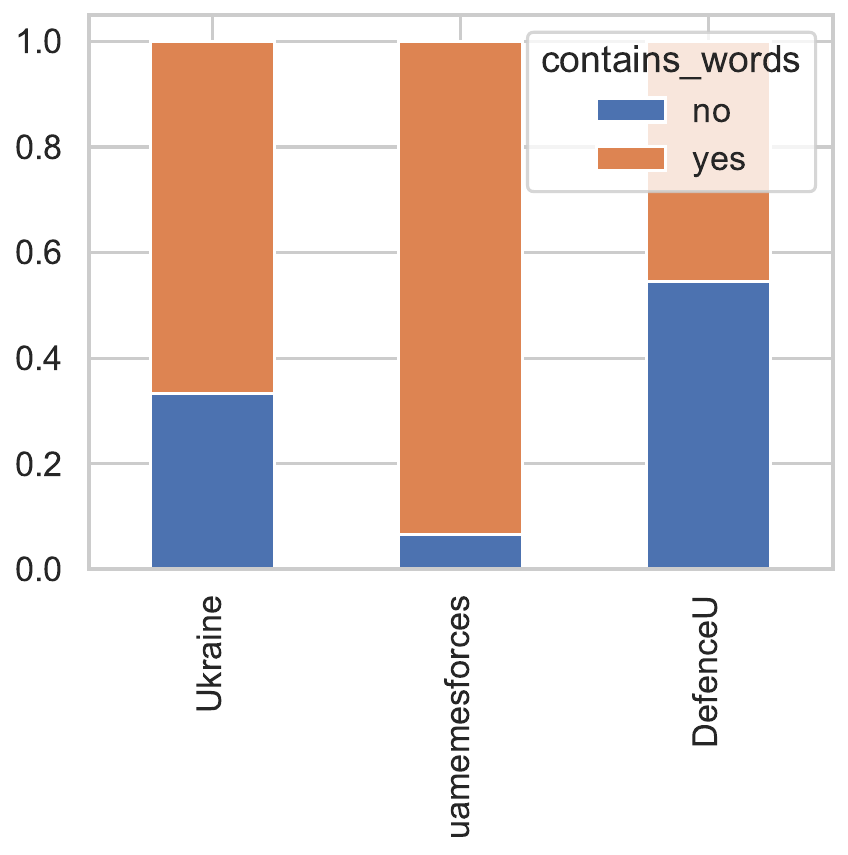}} 
\subfloat[Character emotion]{\includegraphics[width=0.23\textwidth]{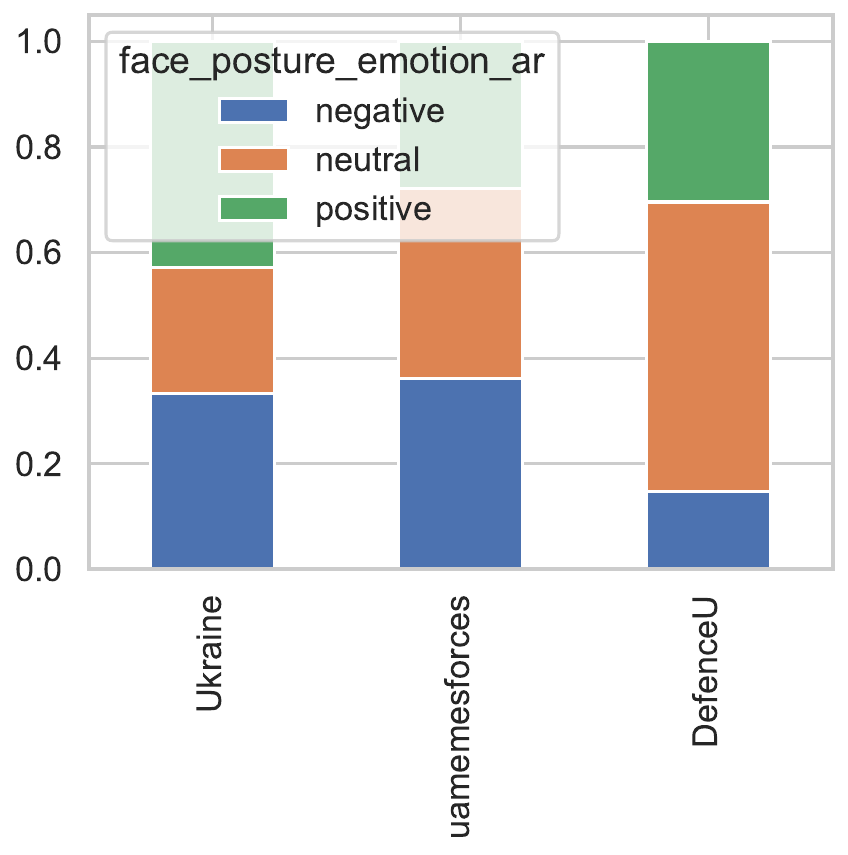}} 
\caption{Distribution of visual features.}
\label{fig:barplots}
\end{figure}

\Cref{fig:barplots} contains summary statistics of content visual features. 
The content produced by the three accounts has markedly different features. 
The \umf tends to publish content spanning both multiple and single panels, have mostly a medium scale, and have a character present, for which both posture and facial expression may be important, and which vast majority of the time contains words.
\ukr and \dou accounts are more similar to each other, in the way that most of their content has a single panel, it uses more illustrations/posters, which contain text.
\dou is especially more likely to have content without words and with neutral emotive expressions.

\end{document}